\documentclass[preprint,12pt]{elsarticle}

\newcommand\blfootnote[1]{%
  \begingroup
  \renewcommand\thefootnote{}\footnote{#1}%
  \addtocounter{footnote}{-1}%
  \endgroup
}

\usepackage{amssymb}
\usepackage{makecell}
\usepackage{amsmath}
\usepackage{url}
\usepackage{subcaption}
\usepackage{multirow}
\usepackage{adjustbox}

\journal{Medical Image Analysis}

\begin{document}

\begin{frontmatter}



\title{A Comprehensive Evaluation of Histopathology Foundation Models for Ovarian Cancer Subtype Classification}


\author[1]{Jack Breen*\footnote[1]{Corresponding author - scjjb@leeds.ac.uk.}}
\author[2,3]{Katie Allen*}
\author[3]{Kieran Zucker}
\author[4]{Lucy Godson}
\author[2,3]{Nicolas M. Orsi\textsuperscript{†}}
\author[1]{Nishant Ravikumar\textsuperscript{†}}
\affiliation[1]{Centre for Computational Imaging and Simulation Technologies in Biomedicine (CISTIB), School of Computing, University of Leeds, UK}
\affiliation[2]{Leeds Institute of Medical Research at St James's, School of Medicine, University of Leeds, UK}
\affiliation[3]{Leeds Cancer Centre, St James's University Hospital, Leeds, UK}
\affiliation[4]{National Pathology Imaging Cooperative (NPIC), Leeds Teaching Hospitals NHS Trust, Leeds, UK}
\blfootnote{*Indicates joint primary authors}
\blfootnote{\textsuperscript{†}Indicates joint senior authors}

\begin{abstract}
Large pretrained transformers are increasingly being developed as generalised `foundation models' which can underpin powerful task-specific artificial intelligence models. Histopathology foundation models show great promise across many tasks, but analyses have typically been limited by arbitrary hyperparameters that were not tuned to the specific task/dataset. We report the most rigorous single-task validation of histopathology foundation models to date, specifically in the context of ovarian cancer morphological subtyping. Attention-based multiple instance learning classifiers were compared using three ImageNet-pretrained feature extractors and fourteen histopathology foundation models. The training set consisted of 1864 whole slide images from 434 ovarian carcinoma cases at Leeds Teaching Hospitals NHS Trust. Five-class classification performance was evaluated using balanced accuracy, AUROC, and F1 scores through five-fold cross-validation, and these cross-validation models were ensembled for hold-out testing and external validation on the Transcanadian Study and OCEAN Challenge datasets. Reporting followed the TRIPOD+AI checklist. The best-performing model used the H-optimus-0 foundation model, with five-class balanced accuracies of 89\%, 97\%, and 74\% in the test sets. 
Normalisations and augmentations aided the performance of the ImageNet-pretrained ResNets, but these were still outperformed by 13 of the 14 foundation models. 
Hyperparameter tuning the downstream classifiers improved performance by a median 1.9\% balanced accuracy, with many improvements being statistically significant.  
Histopathology foundation models offer a clear benefit to ovarian cancer subtyping, improving classification performance to a degree where clinical utility is tangible, albeit with an increased computational burden. Such models could provide a second opinion to histopathologists diagnosing challenging cases and may improve the accuracy, objectivity, and efficiency of pathological diagnoses overall.  Code is made available at \url{https://github.com/scjjb/Ovarian_Features}.

\end{abstract}

\begin{graphicalabstract}
\begin{figure*}[h]
  \centering
\includegraphics[width=\textwidth]{images/GraphicalAbstract.pdf}
\end{figure*}
\end{graphicalabstract}


\end{frontmatter}



\sloppy
\flushbottom

\thispagestyle{empty}

\section*{Introduction}

Ovarian cancer is the eighth most common cancer in women worldwide and typically has a poor prognosis, with 324,000 diagnosed cases translating to 207,000 deaths annually \cite{Bray2024}. It is represented by an array of histological (morphological) subtypes with distinct prognoses and treatment options \cite{Kobel2008}. Five carcinoma subtypes account for approximately 90\% of all ovarian cancers - high-grade serous (HGSC, 70\%), endometrioid (EC, 11\%), clear cell (CCC, 10\%), low-grade serous (LGSC, 5\%), and mucinous carcinomas (MC, 4\%) \cite{Peres2019, Moch2020, Vroobel2024}. 

Histological subtyping is an essential component of the diagnostic process, but it can be challenging. From an individual slide, pathologists only exhibit concordance on an ovarian cancer diagnosis around 80\% of the time \cite{Kobel2014}. In cases of uncertainty, a pathologist may request ancillary tests (such as P53 immunohistochemistry) or seek a second opinion from a gynaecological subspeciality expert, which incurs associated logistical and financial burdens. With increasing cancer rates \cite{Bray2024} and growing complexity in diagnostic testing, histopathology services are increasingly struggling to meet demand globally. For example, most histopathology departments in the UK regularly resort to outsourcing work or hiring temporary staff \cite{RCPath2018}, despite the UK being one of the countries with the most pathologists per capita \cite{Wilson2018}. Any delays resulting from demand outstripping diagnostic resources risk catastrophic impacts on patient outcomes, with a four-week delay in cancer treatment being associated with an approximately 10\% increased mortality rate among patients \cite{Hanna2020}. 

Conceptually, artificial intelligence (AI) may offer clinical value by providing a second opinion to histopathologists, streamlining the diagnostic process and offering additional support when subspecialty experts are not readily available \cite{Allen2024}.  
However, AI models for ovarian cancer diagnosis have yet to demonstrate clinical utility, with most research being small-scale prototyping \cite{Breen2023review} without regulatory approval for clinical use in Europe or the United States \cite{Matthews2024}. AI for ovarian cancer subtyping has constituted a small field of research where, aside from our work \cite{Breen2023, Breen2024ISBI, Breen2024graph}, research has almost exclusively been published by a single group \cite{BenTaieb2015, BenTaieb2016, BenTaieb2017, Levine2020, Boschman2022, Farahani2022, Mirabadi2024, Asadi2024}. While the accuracy of such models has increased over time, the best models still only achieve around 80\% accuracy \cite{Breen2024graph, Farahani2022, Mirabadi2024, Asadi2024, Ma2024,Xu2024} and lack sufficient real-world testing. 

One issue limiting AI in histopathology is that whole slide images (WSIs) are orders of magnitude too large for conventional (single instance) models, therefore multiple instance learning (MIL) is often employed \cite{Gadermayr2024}. In MIL, individual patches (the `instances') are separately processed and then aggregated to learn information about a WSI. These models are impractical to train end-to-end with such large images, so frozen patch feature extractors are often used. As such, any limitation in the pretrained feature extractor can limit downstream classification performance. 

In applying MIL to WSI-level classification, many researchers have used ImageNet-pretrained ResNets \cite{He2016} for patch feature extraction \cite{Breen2023, Boschman2022, Lu2021,Shao2021, Zaffar2023, Godson2024}. ImageNet (a set of 1.4 million natural images from 1000 classes) \cite{Russakovsky2015} is popular for model pretraining as the quantity and diversity of images enables the creation of a multi-purpose feature set. However, these generic features are likely to be suboptimal and computationally inefficient when applied to histopathology images, which contain a relatively homogeneous and restricted set of shapes and colours, with subtle differences being relevant to diagnostic decisions \cite{Vroobel2024,Kobel2010}. 

Recently, many researchers have attempted to create histopathology `foundation models', using self-supervised learning (SSL) techniques to generate broad histopathological feature sets which are not specific to a single organ/cancer type. These approaches have grown rapidly, from tens of thousands of WSIs used to train models with tens of millions of parameters in 2022 and early 2023 \cite{Ciga2022, Chen2022, Wang2022, Kang2023, Filiot2023, Wang2023} to millions of WSIs \cite{Nechaev2024,Vorontsov2024,Zimmermann2024} and billions of parameters more recently \cite{Xu2024,hoptimus0}. 
Foundation models have typically been based on vision transformers (ViTs), utilizing the impressive scalability of transformers seen across many fields, most notably with large language models \cite{Achiam2023, Touvron2023}. Histopathology foundation models have exhibited impressive performance across diverse tasks \cite{Kang2023, Campanella2024, Chen2024,Neidlinger2024} including ovarian cancer subtyping \cite{Breen2024graph,Ma2024,Xu2024}, although analyses have been relatively shallow, without thorough hyperparameter tuning and rigorous statistical comparison of downstream models. Consequently, it is unclear whether models were applied optimally (especially those exhibiting sub-optimal performance), and whether the differences between them were significant. Furthermore, many analyses have been conducted using single-centre data, limiting the assessment of models' generalisability.  



In this study, we present the most comprehensive validation conducted to date comparing feature extraction methods for ovarian cancer subtyping, including three ImageNet-pretrained feature extractors and fourteen histopathology foundation models. 
The analysis includes rigorous hyperparameter tuning and evaluations through five-fold cross-validation, hold-out testing, and external validations, and was conducted with the largest collection of ovarian cancer WSIs used in any AI validation study to date. 
We further investigate whether the classification performance of the ImageNet-pretrained ResNet50 features can match those of the foundation models through stain normalisation, tissue augmentation, or different tissue detection techniques.


\section*{Materials and Methods}
\subsection*{Ovarian Carcinoma Histopathology Data}

\begin{table}[htbp]
\begin{center}
\resizebox{\columnwidth}{!}{%
\begin{tabular}{|c|c|c|c|c|}
\hline
\textbf{\makecell{Carcinoma \\ Subtype}}                     & \textbf{\makecell{Training WSIs \\ (Patients)}}       & \textbf{\makecell{Hold-out WSIs \\ (Patients)}} & \textbf{\makecell{Transcanadian \cite{Kobel2010} \\ WSIs (Patients)}} & \textbf{\makecell{OCEAN \cite{Asadi2024ocean} \\ WSIs}}      \\ \hline 
\makecell{High-Grade Serous \\ (HGSC)}    & 1266 (308)                      & 20 (7)   & 30 (30) &  217                   \\ \hline
\makecell{Low-Grade Serous \\ (LGSC)}     & 92 (21)                         & 20 (6)    & 9 (9)   &  42               \\ \hline
\makecell{Clear Cell \\ (CCC)}            & 198 (45)                       & 20 (7)    & 20 (20)   &   94              \\ \hline
\makecell{Endometrioid \\ (EC)}           &  209 (38)                       & 20 (5)   & 11 (11)    &  119               \\ \hline
\makecell{Mucinous \\ (MC)}               & 99 (22)                        & 20 (5)   & 10 (10)      &   41            \\  \hline
\textbf{Total} & \textbf{1864 (434)} & \textbf{100 (30)} & \textbf{80 (80)} & \textbf{513} \\ \hline 
\end{tabular}%
}
\caption{Dataset breakdown for the training (cross-validation) set, independent internal hold-out test set, and external validation sets. Numbers in brackets indicate the number of unique patients where this is known.}
\label{table:prepdataset}
\end{center}
\end{table}

A training set of 1864 formalin-fixed, paraffin-embedded (FFPE) adnexal tissue WSIs was retrospectively collected from 434 cases of ovarian carcinoma treated at Leeds Teaching Hospitals NHS Trust between 2008 and 2022. Cases were only included if a gynaecological pathologist had diagnosed them as one of the five most common epithelial ovarian cancer subtypes (HGSC, LGSC, CCC, MC, EC). A histopathologist (KA) independently verified all diagnoses, removing any cases with discrepancies. Several representative haematoxylin and eosin (H\&E)-stained adnexal tissue glass slides were selected for each case, cleaned, anonymised, and digitised at 40x magnification using a Leica Aperio AT2 scanner. The population-level class imbalance was reflected in the training set (Table \ref{table:prepdataset}), with the least common subtype (LGSC) represented by only 92 WSIs from 21 cases, compared to 1266 WSIs from 308 cases for the most common subtype (HGSC). 

An independent class-balanced hold-out test set was collected through the same protocol, consisting of 100 primary surgery specimen WSIs from 30 patients. Two additional external test sets were also used. The \emph{Transcanadian Study} dataset \cite{Kobel2010} consisted of 80 WSIs from 80 patients, which had been digitised using an AperioScope scanner and made available at 20x magnification alongside subtype labels that had been determined by a gynaecological pathologist. The \emph{OCEAN Challenge} dataset contained 513 WSIs that had been labelled as one of the five main ovarian carcinoma subtypes. This was a highly heterogeneous dataset, with tissue prepared and digitised across many different labs. However, information was not provided concerning how ground-truth labels were determined and which tissue types were included.  

The main aim of this study was to classify the subtype of primary surgery specimens, which are the initial samples available for pathological analysis. These samples typically have the highest diagnostic quality, as subsequent interval debulking surgery (IDS) samples are impacted by cancer treatments and are thus presumed to have less reliable morphological features. The internal hold-out and Transcanadian Study validation sets contained only primary surgery specimens. It was unclear which samples were included in the OCEAN challenge dataset. The training set contained both primary and IDS specimens, as we have previously found the latter to be beneficial in supplementing training data \cite{Allen2023}.

\subsection*{Slide Classification Pipeline}
\begin{figure*}[h]
  \centering
\includegraphics[width=\textwidth]{images/ABMILpipelineUpdate-compressed.pdf}
\caption{Attention-based multiple instance learning (ABMIL) \cite{Ilse2018} model pipeline for ovarian cancer subtyping, showing the classification of a high-grade serous carcinoma (HGSC).}
\label{fig:abmil}
\end{figure*}

Slide classification was performed using an attention-based multiple instance learning (ABMIL) \cite{Ilse2018} classification pipeline (Figure \ref{fig:abmil}), one of the most commonly used slide classification techniques in contemporary research \cite{Song2023}. WSI preprocessing and patch extractions were performed using the CLAM default procedures \cite{Lu2021}. First, tissue was segmented from plain background using saturation thresholding, where only the pixels with saturation higher than the threshold (8/255) were labelled as tissue. Then, non-overlapping 1024x1024 pixel tissue patches were extracted at the native 40x tissue magnification and downsampled to 256x256 pixels at 10x apparent magnification, which was previously found to be optimal for this task when using the ResNet50 encoder \cite{Breen2024ISBI}. For external data, 512x512 pixel tissue patches were extracted at the native 20x magnification and downsampled to achieve the same 256x256 pixels at 10x apparent magnification. Features were then extracted from these patches following the specific procedure of each feature extraction model, which typically involved first applying a standard normalisation to the red-green-blue (RGB) colour channels, and for ViT-based models typically also involved resizing or cropping patches to 224x224 pixels.


Patch features were then used to train an ABMIL classifier for each feature extractor. In ABMIL, the patch features were passed through a trainable attention layer which assigned each patch an attention score (between 0 and 1) representing the relative importance of the patch in downstream classification. An attention-weighted average of the patch features generated WSI-level features, which were classified through a fully connected neural network with one output node per class. The outputs were passed through the softmax function to generate the (uncalibrated) classification probabilities for each subtype, with the maximum taken as the predicted class. 


\subsection*{Feature Extraction Models}

\begin{table}[h]
    \centering
    
    \begin{adjustbox}{width=1.4\textwidth,center}
    \begin{tabular}{|c|cccccccc|}
         \hline
         \textbf{\makecell{Feature \\ Extractor}}&  \textbf{Backbone}&  \textbf{\makecell{Data \\ Type}}&  \textbf{Data Source} &\textbf{\makecell{Pretraining \\ Algorithm}}&  \textbf{\makecell{Pretraining \\ Images}}& \textbf{\makecell{Pretraining \\ Magnification(s)}} &   \textbf{Parameters}&  \textbf{\makecell{Patch \\ Features}} \\
         \hline
         RN50 \cite{He2016}&  ResNet50& Natural  & ImageNet-1k &  Supervised& 1,431,167 & NA & 8,543,296 &  1024 \\
         RN18 \cite{He2016}&  ResNet18& Natural  & ImageNet-1k & Supervised& 1,431,167 & NA & 11,176,512  &  512 \\
         ViT-L \cite{Dosovitskiy2020}&  ViT-L&  Natural& ImageNet-21k &  Supervised& 14,197,122 & NA & 303,301,632 &  1024 \\
         \hline
        RN18-Histo \cite{Ciga2022}& ResNet18& Histo& 57 Online Sets & SimCLR & $>$25,000 WSIs & 10x,20x,40x,100x & 11,176,512 & 512 \\ 
        Lunit \cite{Kang2023} & ViT-S & Histo & TCGA + Internal & DINO & 36,666 WSIs & 20x,40x & 21,670,272 & 384\\ 
        RN50-Histo \cite{Kang2023} & ResNet50 & Histo & TCGA + Internal & Barlow Twins & 36,666 WSIs & 20x,40x & 23,508,032 & 2048\\ 
         CTransPath \cite{Wang2022}&  CNN + SwinT &  Histo& TCGA + PAIP & Novel SSL & 32,220 WSIs & 20x & 27,520,038&  768\\
        Hibou-B \cite{Nechaev2024}&  ViT-B&  Histo& Internal & DINOv2 & 1,141,581 WSIs & Unclear & 85,741,056&  768
 \\
         Phikon \cite{Filiot2023}&  ViT-B&  Histo& TCGA& iBOT& 6,093 WSIs & 20x & 85,798,656&  768
 \\
         Kaiko-B8 \cite{Aben2024}&  ViT-B&  Histo& TCGA& DINO& $\sim$29,000 WSIs  & 5x,10x,20x,40x & 85,807,872 &  768
 \\
         GPFM \cite{Ma2024} &  ViT-L&  Histo& 47 Online Sets & Novel Distillation& 72,280 WSIs & Unclear  & 303,228,928 &  1024
 \\
         UNI \cite{Chen2024}&  ViT-L&  Histo& Internal + GTEx & DINOv2& 100,426 WSIs & 20x  & 303,350,784 &  1024
 \\      
         Hibou-L \cite{Nechaev2024}&  ViT-L&  Histo& Internal & DINOv2& 1,141,581 WSIs & Unclear  & 303,659,264 &  1024 \\      
         Virchow \cite{Vorontsov2024}&  ViT-H&  Histo& Internal & DINOv2 & 1,488,550 WSIs & 20x &  631,229,184&  2560
 \\
          Virchow2-CLS \cite{Zimmermann2024} &  ViT-H&  Histo& Internal & DINOv2 & 3,134,922 WSIs & 5x,10x,20x,40x & 631,239,424
 &  1280
 \\
         H-optimus-0 \cite{hoptimus0} &  ViT-g&  Histo& Internal &  DINOv2& $>$500,000 WSIs & 20x & 1,134,774,272 &  1536
 \\
         Prov-GigaPath \cite{Xu2024}&  ViT-g&  Histo& Internal &  DINOv2& 171,189 WSIs & 20x &  1,134,953,984&  1536
 \\ \hline
    \end{tabular}
    \end{adjustbox}
    \caption{Summary of the seventeen feature extraction models.}
    \label{tab:featureextractors}
\end{table} 

A total of seventeen patch feature extractors were compared (Table \ref{tab:featureextractors}), three of which had been trained through the traditional approach of supervised classification on ImageNet data \cite{Russakovsky2015}, and the other fourteen had been trained using histopathology images through various self-supervised learning (SSL) approaches. All feature extractors were available online, with some requiring approval before they could be accessed.

The ImageNet-pretrained models were a ResNet50 \cite{He2016}, ResNet18 \cite{He2016}, and a large vision transformer (ViT-L) \cite{Dosovitskiy2020}. The ResNet50 outputs were taken from the end of the third residual block (as in CLAM \cite{Lu2021}) to give 1024 features per input patch. The ResNet18 does not have a layer this large, so 512 features were extracted from the end of the fourth residual block instead. ViT-L was applied without a final fully connected layer to give 1024 features per patch. 
ImageNet-pretraining for ResNet models had been conducted using the original 1,000 class ImageNet dataset alone, whereas the ViT-L was first trained on the much larger set of nearly 22,000 classes, and then fine-tuned to the same set of 1,000 classes. The reported ImageNet classification accuracies were 80.9\%, 69.8\%, and 85.1\% for ResNet50, ResNet18, and ViT-L, respectively.

The SSL pretraining of the foundation models allowed large quantities of diverse data to be leveraged 
without the need for extensive labelling. 
One of the earliest histopathology foundation models was a ResNet18 trained through a self-supervised strategy with 57 online datasets in 2021 \cite{Ciga2022}, which we refer to as `RN18-Histo'. A similar approach was taken in a subsequent study to pre-train a ResNet50 with a combination of TCGA and proprietary data, which we refer to as `RN50-Histo' \cite{Kang2023}.
Another early approach, CTransPath \cite{Wang2022}, used a novel backbone which combined a CNN with a Swin Transformer, and pretrained these through a novel SSL strategy using multiple online datasets.  

Newer histopathology foundation models have typically used vision transformer backbones. The smallest such model, Lunit \cite{Kang2023}, was based on the small vision transformer backbone (ViT-S), which gave a model of a similar size as RN50-Histo that had been pretrained on the same dataset (using DINO). 
Three of the foundation models were built using the base vision transformer (ViT-B) backbone with different pretraining procedures, with Phikon \cite{Filiot2023} trained using iBOT on a small subset of TCGA data, Kaiko-B8 \cite{Aben2024} on a much larger set of TCGA data using DINO, and Hibou-B \cite{Nechaev2024} on a huge proprietary dataset using DINOv2. The authors of Kaiko-B8 also made their model available with four other backbone sizes, though the B8 variation gave the best overall performance in their evaluations \cite{Aben2024}. 
Hibou-B was included as it was the best-available version of this model when initial validations were conducted, although the authors reported their larger model, Hibou-L, to have given better performance \cite{Nechaev2024}.   

The largest histopathology foundation models (all published in 2024) have typically trained larger vision transformers with proprietary datasets of over 50,000 WSIs using DINOv2 \cite{Oquab2023}. GPFM \cite{Ma2024}, UNI \cite{Chen2024}, and Hibou-L are large vision transformers (ViT-L) trained with 72,280 WSIs, 100,426 WSIs, and 1,141,581 WSIs, respectively. 
Virchow \cite{Vorontsov2024} and its recent update, Virchow2 \cite{Zimmermann2024}, are huge vision transformers (ViT-H) trained with the largest dataset for any histopathology foundation model to date, with nearly 1.5m WSIs in the first version and over 3m WSIs in the second version. Virchow also has the largest feature space as the class tokens are concatenated with the average patch tokens, where typically only the class tokens would be used. As Virchow2 was reported by the original authors to give better results using just the class tokens \cite{Zimmermann2024}, we adopted this version as `Virchow2-CLS'. 

Prov-GigaPath \cite{Xu2024} and H-optimus-0 \cite{hoptimus0} were the largest accessible histopathology foundation models by far, with the ViT-g backbone giving over one billion parameters, nearly twice as many as the next largest (Virchow2-CLS), and over 100x as many parameters as the smallest foundation model (RN18-Histo). These models had also been trained with hundreds of thousands of WSIs. Prov-GigaPath includes a patch-to-slide aggregator, though we focused only on the patch feature extractor.

\subsection*{Normalisation and Augmentation Analysis}
To investigate whether the baseline ImageNet-pretrained ResNet50 encoder could be made competitive with the modern alternatives, we applied this feature extractor with a variety of data preprocessing techniques. Seven approaches were evaluated, with two applying stain normalisations (Reinhard \cite{Reinhard2001} and Macenko \cite{Macenko2009}), two applying Otsu thresholding \cite{Otsu1975} for tissue detection (with and without Macenko normalisation), and three applying data augmentation (increasing the apparent dataset size by factors of 5, 10, and 20). These methods are described in further detail in \ref{app:augmentation}.

\subsection*{Hyperparameter Tuning and Evaluation Procedures}

ABMIL classifiers were tuned using an iterative grid search where typically two hyperparameters were adjusted at a time, with the best taken forward to the next iteration. Ten hyperparameters were tuned using the average loss of the five-fold validation sets. Seven of these hyperparameters directly influenced the Adam optimiser \cite{Kingma2014}, controlling the learning rate, learning rate decay proportion and patience, first and second moment decay, optimisation stability, and L2 regularisation rate. The remaining hyperparameters controlled the model size (the dimension of the attention layer and subsequent fully connected layer), and the proportions of parameter dropout and data dropout during training. Models were trained using a balanced cross-entropy loss and class-weighted sampling to help account for the class imbalance in the training set. Initial hyperparameters were determined based on a previous study in which ABMIL was tuned using ResNet50 features for the same task with a smaller dataset \cite{Breen2024ISBI}. Over 150 unique hyperparameter configurations were evaluated during the tuning of each classifier.

Models were evaluated using the balanced accuracy, macro-averaged area under the receiver operating characteristic curve (AUROC), and macro F1 score. These metrics assessed different aspects of classification performance, with AUROC giving a holistic but imbalanced overview of discriminative power, F1 giving a balanced measure of predictive performance at a specific threshold, and balanced accuracy representing realistic clinical performance. Stratified five-fold cross-validation (split 60-20-20 train-val-test at the case level to avoid data leakage) was employed during training. 
In hold-out testing and external validations, the predictions of the five cross-validation models were averaged to generate an ensembled classification. All results were reported using the mean and 95\% confidence intervals from 10,000 iterations of bootstrapping. An ablation study was also conducted to investigate whether hyperparameter tuning improved model performance, with the performance of the tuned models compared to those using the default hyperparameters.

Paired t-tests were used to test for statistically significant differences between each model and the baseline ResNet50 across the five cross-validation folds, with p-values adjusted for multiple testing using a false discovery rate correction \cite{Benjamini1995}. Results were considered \emph{statistically significant} given an adjusted p-value $<$ 0.05. Paired t-tests were also used in the hyperparameter tuning ablation to determine whether tuning the ABMIL classifiers had a statistically significant effect on the final results. 

This manuscript was prepared following the TRIPOD+AI (Transparent Reporting of a multivariable prediction model for Individual Prognosis Or Diagnosis + Artificial Intelligence) checklist \cite{Collins2024} to ensure thorough reporting, with the completed checklist available in \ref{app:tripod}. The PyTorch-based code used in this study is available at \url{https://github.com/scjjb/Ovarian_Features}. Experiments were conducted using an NVIDIA A100 GPU and 32 AMD EPYC7742 CPUs @3.4GHz.

\section*{Results}

\begin{figure}[htbp]
\centering
\includegraphics[width=\textwidth]{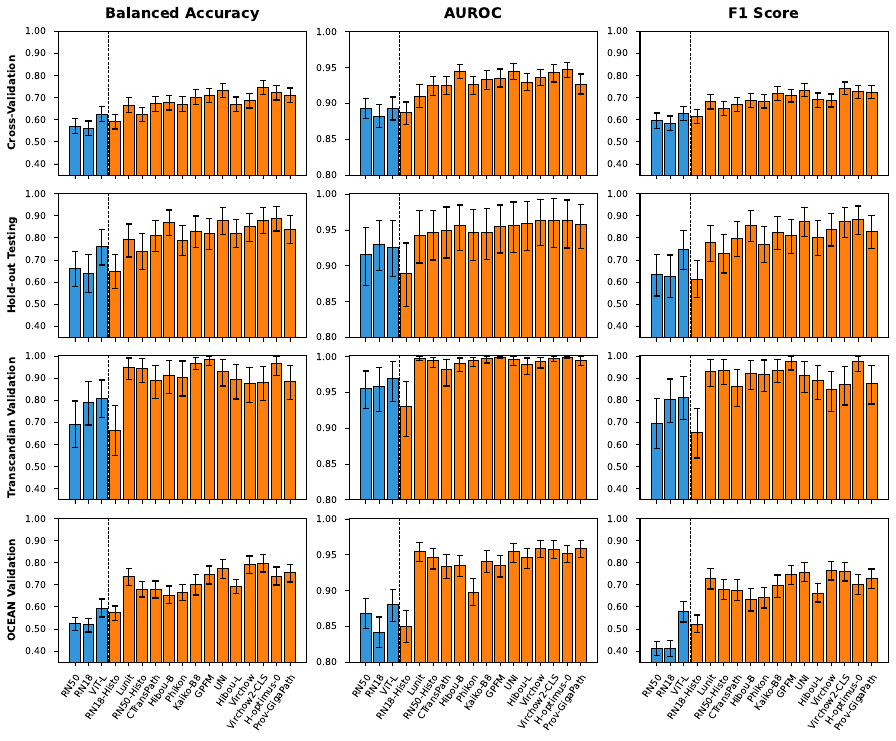}
\caption{Ovarian cancer subtyping results for each feature extractor (mean and 95\% confidence interval generated by 10,000 iterations of bootstrapping). Blue indicates ImageNet-pretrained feature extractors, orange indicates histopathology foundation models. Hold-out testing and external validation results are based on an ensemble of cross-validation models. Precise values are tabulated in \ref{app:results}.}
\label{fig:results}
\end{figure}

 As shown in Figure \ref{fig:results}, no single model gave the greatest results in every validation, with Virchow2-CLS giving the greatest performance in cross-validation, H-optimus-0 in hold-out testing and the Transcanadian Study external validation, and Virchow in the OCEAN Challenge external validation. RN18-Histo had the worst performance of any foundation model in all validations and was the only foundation model to perform worse than an ImageNet-pretrained encoder overall. Expanded results are provided in \ref{app:results} and \ref{app:stats}.

\begin{table}[h]
    \centering
    \begin{tabular}{|c|c|ccc|}
         \cline{2-5}
         \multicolumn{1}{c|}{} & \textbf{\makecell{Feature \\ Extractor}}& \textbf{\makecell{Balanced\\Accuracy}}   
&  \hspace{0.5em}\textbf{AUROC}\hspace{0.5em} & \textbf{F1 Score}
\\
         \hline
         \multirow{3}{*}{\centering \makecell{ImageNet- \\ Pretrained \\ Models}} & RN50& 61.2\%   
& 0.908 &  0.585 
\\
       &  RN18& 62.8\%   
& 0.903 &  0.607  
\\
       &  ViT-L& 69.7\%   
& 0.917 &  0.692  
\\
\hline
       \multirow{14}{*}{\centering \makecell{Histopathology \\ Foundation \\ Models}} &  RN18-Histo& 62.0\%   
& 0.889 & 0.601   
\\
        & Lunit& 78.6\%   
 & 0.951 & 0.780
\\
       &  RN50-Histo& 74.7\%   
& 0.953 &  0.749  
\\
       &  CTransPath& 76.2\%  
& 0.948 & 0.751  
\\
       &  Hibou-B& 77.9\%   
& 0.957 & 0.775   
\\
       &  Phikon& 75.7\%   
& 0.941 &  0.754  
\\
       &  Kaiko-B8& 80.0\%  
& 0.955 &  0.794  
\\
        & GPFM& 81.4\% & 0.956 & 0.811 \\
       &  UNI& 82.9\%   
& 0.963 &  0.820  
\\
        & Hibou-L& 76.9\%   
& 0.956 &  0.762  
\\
        & Virchow& 80.1\%   
& 0.963 &  0.785 
\\
       &  Virchow2-CLS& 82.6\%   
& \textbf{0.966} &  0.811 
\\
       &  H-optimus-0& \textbf{83.0\%}   
& 0.965 & \textbf{0.822}  
\\
        & Prov-GigaPath& 79.8\%   
& 0.960 & 0.791   
\\
 \hline
    \end{tabular}
    \caption{Averaged results across the four validations. The greatest result for each metric is shown in \textbf{bold}.}
 \label{table:avgresults}
\end{table}

\begin{figure}[h]
\centering
\includegraphics[width=0.95\textwidth]{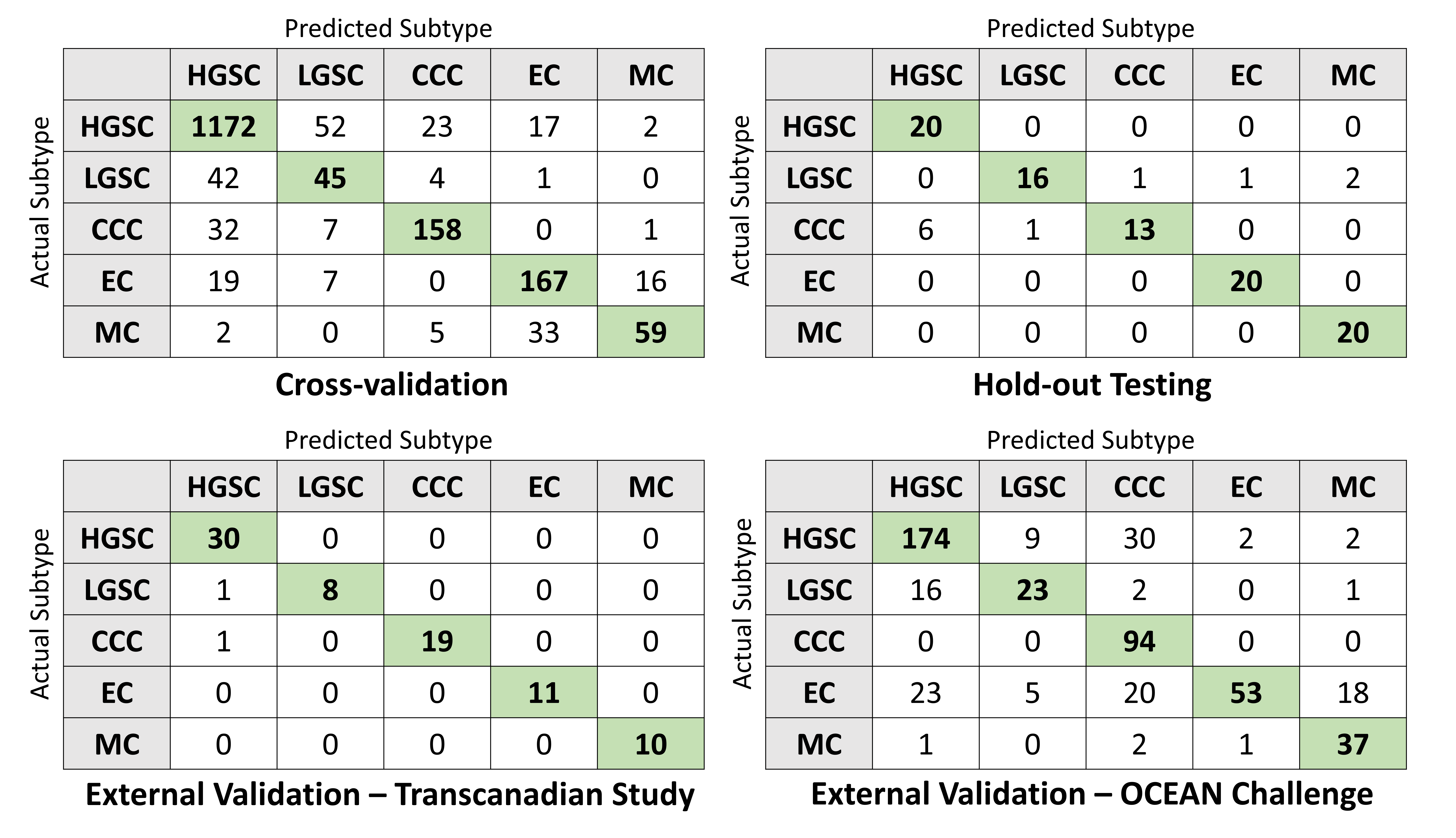}
\caption{Confusion matrices for the optimal ABMIL classifier with features from the H-optimus-0 foundation model. Correct classifications are indicated in green.}
\label{fig:confusion}
\end{figure}

The H-optimus-0 model achieved the greatest averaged performance across all validations (Table \ref{table:avgresults}), with 83.0\% average balanced accuracy, 0.965 average AUROC, and 0.822 average F1 score. 
This performance was closely followed by that of UNI and Virchow2-CLS. The worst averaged performances were given by CNN-based feature extraction models (RN50, RN18, RN18-Histo), followed by the ImageNet-pretrained vision transformer. Confusion matrices for the optimal H-optimus-0 model are provided in Figure \ref{fig:confusion}. 

The difference in performance between each foundation model (except RN18-Histo) and the baseline Imagenet-pretrained ResNet50 was found to be significant by all metrics in all validations, except by the AUROC in cross-validation (for nine foundation models), RN50-Histo by most metrics in internal validations, and Hibou-B by balanced accuracy in the external validation on the OCEAN Challenge dataset. There was no significant difference between the performance of the baseline model and either the RN18 feature extractor or the RN18-Histo foundation model in most validations. The difference between the baseline ResNet50 and the ViT-L feature extractor was statistically significant in most validations for the balanced accuracy and F1 score, but not the AUROC. The p-values are tabulated in \ref{app:stats}. 


\clearpage

\subsection*{Normalisation and Augmentation Results}

\begin{figure}[h]
\centering
\includegraphics[width=0.95\textwidth]{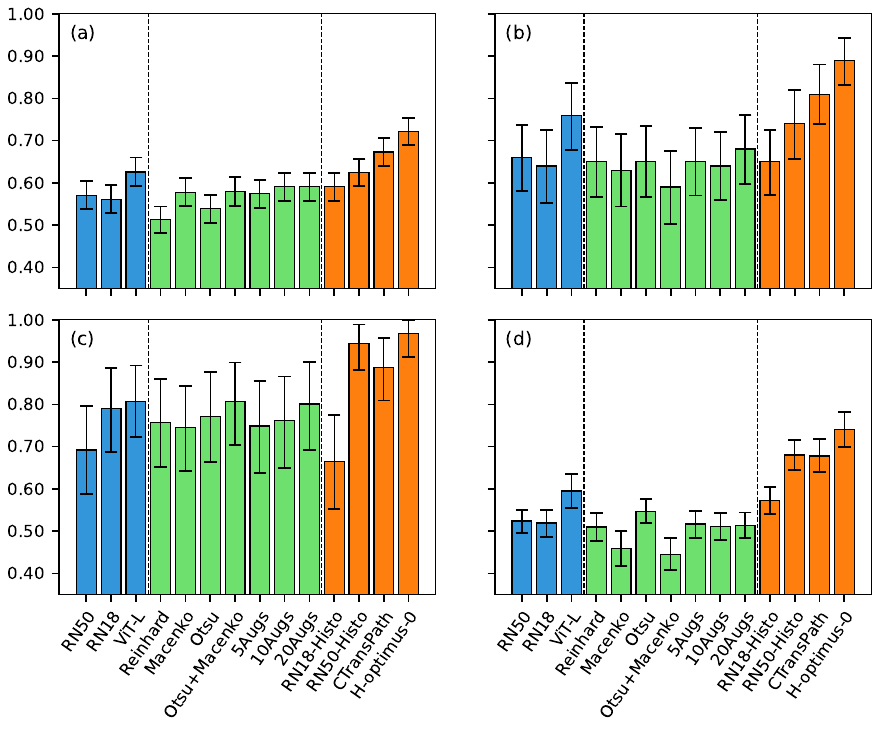}
\caption{Balanced accuracy results for each ImageNet-pretrained feature extractor (blue), including the seven ResNet50 models with varied preprocessing techniques (green), as well as the three worst-performing (RN18-Histo, RN50-Histo, and CTransPath) and the best-performing foundation models (H-optimus-0) in (a) cross-validation, (b) hold-out testing, (c) external validation on the Transcanadian Study dataset, (d) external validation on the OCEAN Challenge dataset. For validations (b)-(d), predictions were ensembled from the five cross-validation models. Results reported as the mean and 95\% confidence interval generated by 10,000 iterations of bootstrapping. Precise values and other metric results are tabulated in \ref{app:augmentation}.}
\label{fig:results_norm}
\end{figure}

As shown in Figure \ref{fig:results_norm}, different preprocessing techniques had inconsistent effects on the baseline ResNet50 feature extractor, with modest benefits in internal validations, and variable effects in external validations. In cross-validation, 
no pre-processing method improved the balanced accuracy or F1 score by more than 0.02, and improvement was seen in AUROC with any method. 
In hold-out testing, 
only the 20x colour augmentation improved performance, increasing F1 by 0.023 and balanced accuracy by 0.020, but reducing AUROC by 0.012.
However, in the external validation on the Transcanadian Study dataset, every preprocessing method improved performance compared to the baseline by over 0.05 balanced accuracy and F1 score, and over 0.002 AUROC.  
The greatest performances in this validation were found by combining Otsu thresholding with Macenko normalisation and by 20x colour augmentations, which each increased the F1 score and balanced accuracy above baseline performance by over 0.1, and AUROC by over 0.016. 
For the OCEAN Challenge dataset, most preprocessing methods gave worse results than the baseline approach, with only Otsu thresholding providing any benefit over the baseline performance. 

Despite some modest improvements offered by different preprocessing techniques, particularly in the Transcanadian Study external validation, the best-performing model based on the ImageNet-pretrained ResNet50 backbone was still outperformed by every foundation model (except RN18-Histo) in every validation. Furthermore, none of the different preprocessing methods gave statistically significant differences in performance compared to the baseline approach in any validation.  These results are explored further in \ref{app:augmentation}.

\clearpage
\subsection*{Hyperparameter Tuning Ablation Results}

\begin{figure}[htb]
\centering
\includegraphics[width=0.99\textwidth]{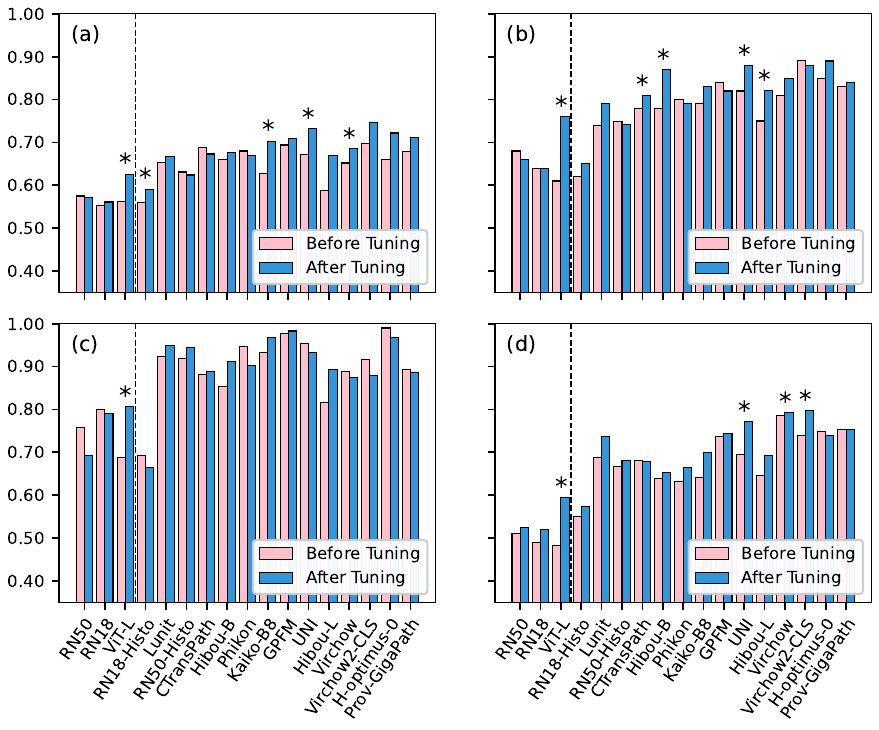}
\caption{Balanced accuracy results for each model compared with the ABMIL classifier trained using the default hyperparameters (pink) and the tuned hyperparameters (blue) for (a) cross-validation, (b) hold-out testing, (c) external validation on the Transcanadian Study dataset, (d) external validation on the OCEAN Challenge dataset. For validations (b)-(d), predictions were ensembled from the five cross-validation models. *Indicates a significant difference in the paired t-test at the 5\% significance level.
}
\label{fig:tuningresults}
\end{figure}

The balanced accuracy results of each ABMIL classifier trained with the default hyperparameters (without tuning) are compared to the tuned model results in Figure \ref{fig:tuningresults}, with exact values and other metrics provided in \ref{app:tuningablation}. The median impact of hyperparameter tuning was an improvement of 1.9\% balanced accuracy, 0.005 AUROC, and 0.025 F1 score, though the effect on any given model in any given validation was variable, with balanced accuracies changed by $-$6.6\% to $+$15.0\%, AUROCs by $-$0.013 to $+$0.041, and F1 scores by $-$0.073 to $+$0.146. The only models which did not benefit from hyperparameter tuning were those using the ResNet50, ResNet18, Phikon, and H-optimus-0 feature extractors. All of the other models had a statistically significant difference between tuned and untuned results in at least one evaluation (\ref{app:stats}), with these significant differences only occurring in cases where tuning improved performance. The optimal hyperparameters from tuning each model and the loss from each hyperparameter tuning iteration are reported in \ref{app:hyperparams}.

\clearpage
\section*{Discussion}

In this study, we thoroughly compared the effects of different patch feature extractors on the slide-level classification of ovarian carcinoma morphological subtypes. The results indicated that transformer-based histopathology foundation models improved downstream classification when compared to non-domain-specific and ResNet-based feature extractors, with 13 out of 14 foundation models outperforming all ImageNet-pretrained models in all evaluations.  The strong performance of foundation models was particularly impressive considering that they were applied here at 10x magnification, despite often only being trained using 20x magnification data.

The only foundation model which did not exceed ImageNet-pretrained model performance was RN18-Histo, which was the single worst-performing model in hold-out testing and external validation on the Transcanadian Study, though it did outperform the ImageNet-pretrained ResNet models in the other two validations. RN18-Histo was the earliest published histopathology foundation model and as such it was one of the few foundation models to not use a transformer-based backbone. In this study, RN18-Histo was also the smallest foundation model, had the second-smallest feature space, and was pretrained with the second-smallest dataset.  

\begin{figure}[h]
\centering
\includegraphics[width=\textwidth]{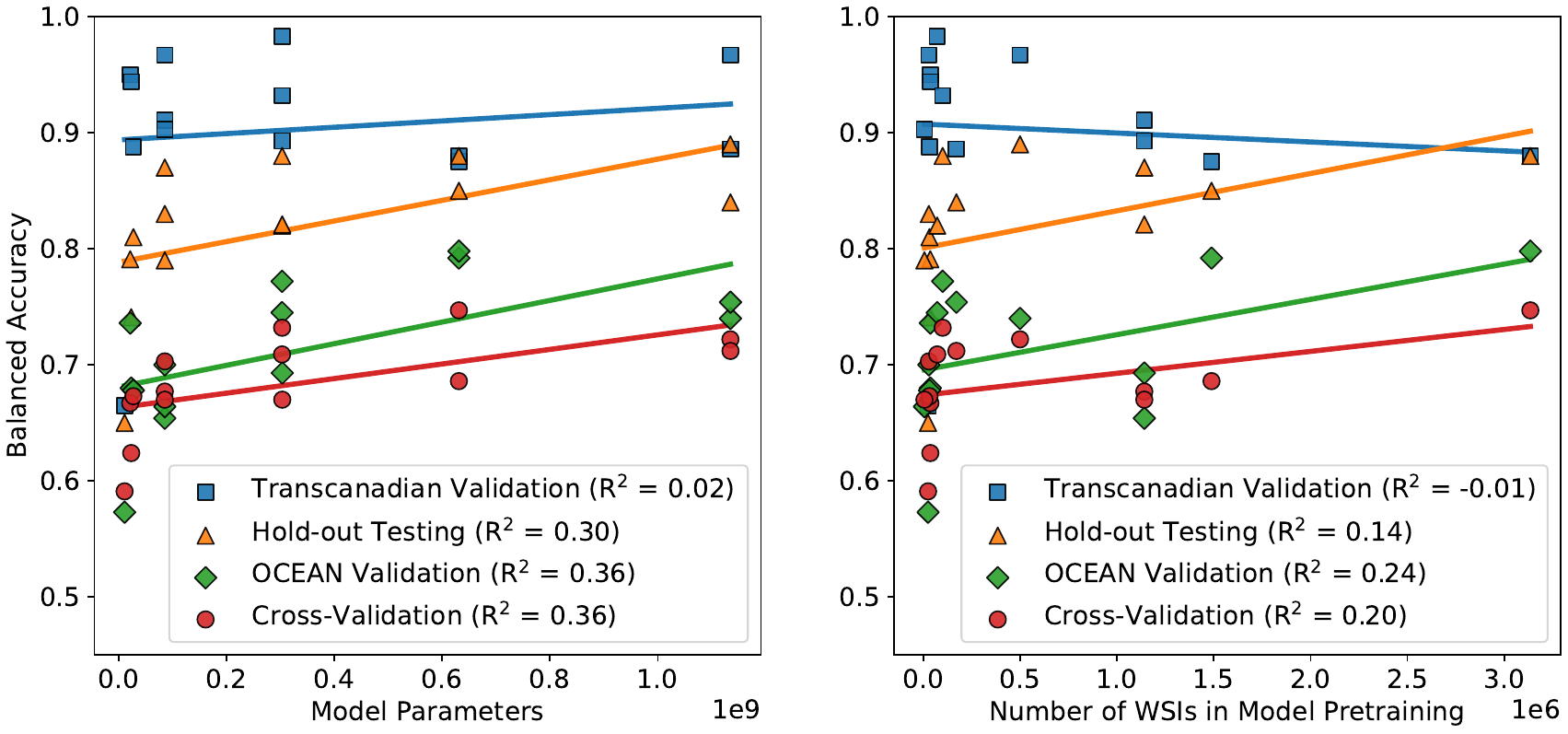}
\caption{Balanced accuracy results for each histopathology foundation model-based classifier in each validation shown in relation to the number of model parameters and number of WSIs used in the pretraining of the foundation model. The line of best fit and the corresponding coefficient of determination (R$^2$) are provided for each validation.
}
\label{fig:modelsize}
\end{figure}

As shown in Figure \ref{fig:modelsize}, in most validations there was a slight positive relationship between performance (specifically, balanced accuracy) and each of the foundation model size and pretraining dataset size. These relationships were fairly weak, with the relationship between performance and foundation model size having R$^2$ values between 0.02 and 0.36, and the relationship between performance and pretraining dataset size between -0.01 and 0.24. 
The greatest performance in most validations was achieved by one of the largest models (Virchow, Virchow2-CLS, H-optimus-0), though in the Transcanadian Study external validation the smaller GPFM model performed best, and the single largest model (Prov-GigaPath) did not achieve optimal results in any validation. Three models were trained with over one million WSIs, with two being among the best-performing models (Virchow, Virchow2-CLS), and the other being one of the worst-performing ViT-based foundation models overall (Hibou-B).

To investigate which foundation models outperformed expectations, we investigated which models had positive residuals of at least 1\% when compared to the lines of best fit in Figure \ref{fig:modelsize}. 
UNI and Kaiko-B8 consistently performed better than expected given their foundation model size, with GPFM and Virchow2-CLS performing better than expected in three of four validations. The UNI and GPFM models consistently performed better than expected given the pretraining dataset size, with Kaiko-B8, Virchow2-CLS and H-optimus-0 all better than expected in three of four validations. These results indicate that UNI is particularly data-efficient and computationally-efficient for a foundation model of its ability. It is not clear how UNI outperformed expectations in this way, with similar overall methodologies employed in training models which did not achieve such great results. The proportion of gynaecological WSIs in the UNI training set (5.8\%) was exceeded in the training of several other models \cite{Ma2024,Ciga2022,Filiot2023,Nechaev2024,Vorontsov2024}, though for most models it was not clear what proportion of the training set was specifically composed of the five subtypes of interest, so it is not clear whether this was an influential factor.   

Different preprocessing techniques often had little impact on internal performance (likely due to the homogeneity of the single-centre dataset) and on the OCEAN Challenge validation set, but they did aid the generalisability to the Transcanadian Study dataset.  There was a modest positive trend between the number of augmentations used and the resulting model performance which may continue beyond the x20 augmentations used herein, though this may not be worth the considerable associated computational burden since the normalisation approaches achieved a similar level of performance. No individual normalisation, augmentation, or tissue detection approach consistently improved performance, with each giving worse performance than the baseline in at least one validation. As such, we believe there is much greater value in selecting the optimal feature extractor than there is in applying varied preprocessing techniques in the training of a downstream classifier.

Performance was generally higher in hold-out testing than in cross-validation and was higher still in the external validation with the Transcanadian Study dataset. However, the external validation with the OCEAN dataset gave similar performance to that of cross-validation. This may be influenced by the diagnostic quality of the data - validations using only staging data achieved optimal balanced accuracies of 89\% and 97\%, compared to only 75\% and 80\% in the validations including IDS samples (which can pose diagnostic challenges due to chemotherapy-induced morphological changes, such as varying amounts of cell death and associated changes in surrounding stroma). The Transcanadian Study set contained a single representative staging slide of the tumour per patient and the slides were largely devoid of artifacts. This particularly high-quality data may represent a best-case research scenario, rather than a more realistic representation of the variable quality and tumour content of clinical slides, where guidance recommends the sampling of heterogeneous areas of tumour that have the potential to compromise the quality of slide preparation and interpretation, such as calcification or necrosis. The hold-out and external validations likely also benefitted from the five-fold ensembled predictions when compared to the five-fold cross-validation.  While this is the most comprehensive study of AI ovarian cancer subtyping to date, the relatively small size of the test sets still results in a high level of uncertainty, as reflected by the wide confidence intervals. Thus, part of the difference in performance between datasets may be attributed to random chance. 

The results of this study are similar to those of the only previous studies to use large ovarian cancer subtyping datasets (each with around 1000 WSIs) \cite{Farahani2022, Mirabadi2024, Asadi2024}. One study presented a multi-scale graph model \cite{Mirabadi2024} and reported an optimal cross-validation balanced accuracy of 73\% and F1 score of 0.69
, respectively.
Another \cite{Farahani2022} evaluated four MIL approaches and reported an optimal cross-validation balanced accuracy of 81\%, AUROC of 0.95, and  F1 score of 0.79
. In an external validation using an ensemble of cross-validation models on 60 WSIs, the authors reported a balanced accuracy of 80\%, AUROC of 0.96, and F1 score of 0.81
. The final study focused on adversarial domain adaptation \cite{Asadi2024} and achieved optimal internal and external balanced accuracies of 80\% and 83\% from a CTransPath-based MIL classifier. 
Other studies applying foundation models to ovarian cancer subtyping have reported optimal balanced accuracies of 82\% and $\sim$88\% using UNI on the OCEAN dataset and Prov-GigaPath on an internal dataset, respectively \cite{Ma2024,Xu2024}. 
These comparisons are provided for context and should not be considered to be conclusive given the differences in the datasets used. 
A sparsity of publicly available data has limited external validations in most previous research \cite{Breen2023review}, and for the largest accessible dataset (the OCEAN Challenge set) very little information has been provided about the data provenance.    

\begin{figure}[htbp]
\centering
\includegraphics[width=0.95\textwidth]{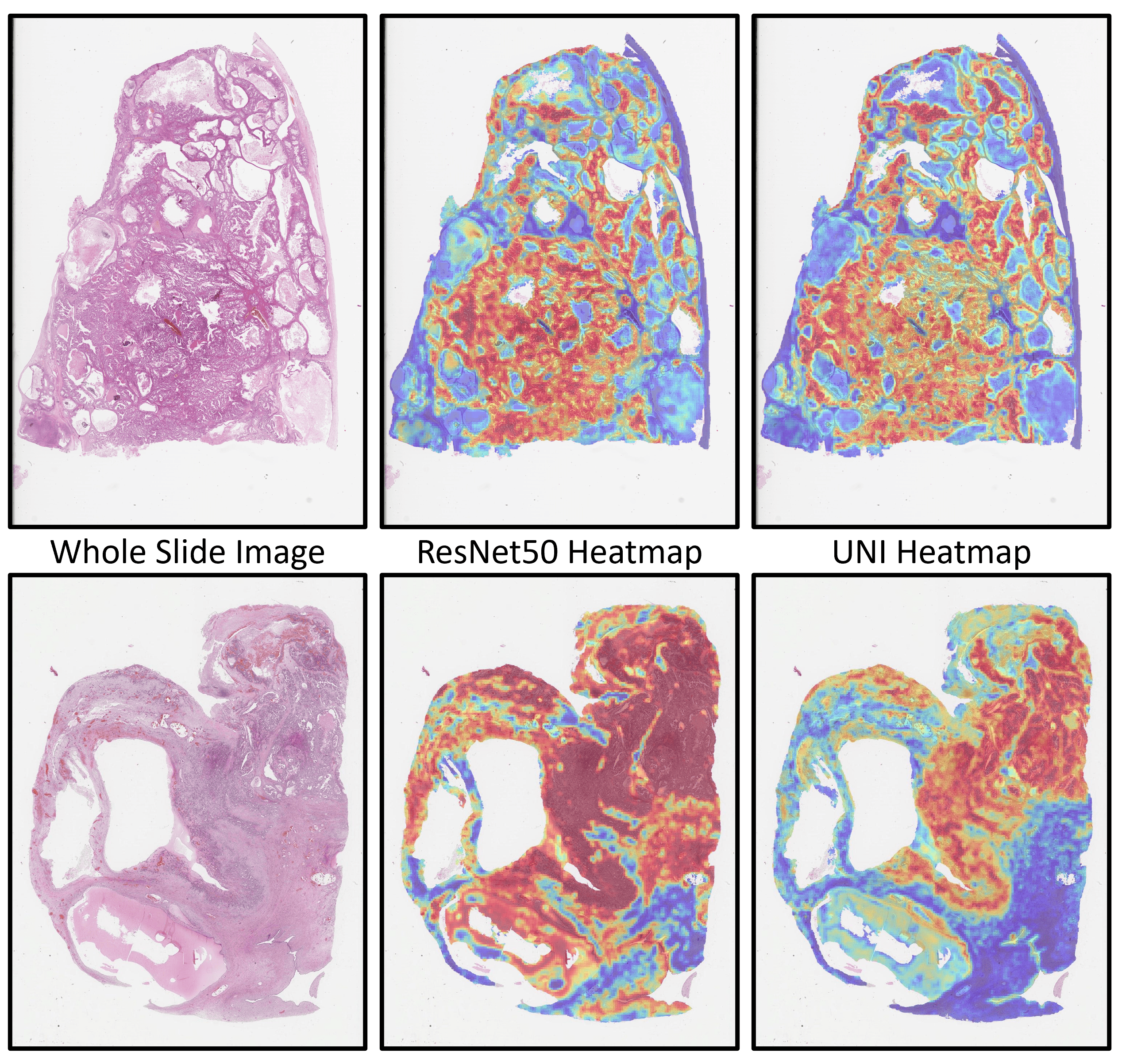}
\caption{Attention heatmaps from the ABMIL classifier using the ImageNet-pretrained ResNet50 and histopathology-UNI foundation model features. (Upper) A typical difference between heatmaps with different diagnoses. (Lower) The most extreme qualitative difference found between heatmaps in the internal test set. In both examples, the UNI classification was correct (upper - MC, lower - CCC) , and the ResNet50 classification was incorrect (upper - EC, lower - MC).
These heatmaps are based on 256x256 pixel patches with 50\% overlap at 10x apparent magnification, with visual differences caused by the variable size of resection samples.
}
\label{fig:heatmaps}
\end{figure}

To qualitatively analyse the differences between foundation models and ImageNet-pretrained CNNs, two pathologists (KA and NMO) qualitatively compared the ABMIL attention heatmaps (Figure \ref{fig:heatmaps}) generated using the baseline ResNet50 and the UNI foundation model. 
Most heatmaps were well-focused on tumour and relevant stromal regions for both models, with often only subtle differences between them. The UNI-based heatmaps generally indicated a slightly greater focus on tumour tissue, whereas the ResNet50 model also paid attention to some stromal regions of variable diagnostic relevance (\ref{app:heatmaps}). Attention heatmaps can be useful for identifying potential sources of error but should be interpreted with caution since they cannot provide a complete explanation of classification decisions \cite{Bibal2022}.  



In this study, we reported the second-highest ever performance of an AI model for ovarian cancer subtyping (behind our concurrent study using multi-resolution graph networks with the UNI encoder), with 97\% balanced accuracy on the Transcanadian Study dataset. 
However, results were variable across datasets.  
The improved performance from histopathology foundation models is promising for the potential clinical utility of these AI approaches, though further work is required to ensure that the models generalise to all relevant sources of variation, especially across different histopathology labs and slide scanners. This should include robustness to lower quality data and artifacts to reduce the burden of quality control, which otherwise risks sacrificing any time savings the models could provide to diagnostic services.
Furthermore, it is currently unclear how best to present automatically generated information to pathologists to assist them, rather than to distract, frustrate, or confuse them. This may require improved model interpretability and a measure of model uncertainty, especially considering the existence of rare subtypes which are notoriously difficult to collect sufficient data on outside the context of multi-centre collections. 


Ideally, algorithms would be made more computationally efficient for use in the clinic, 
but the best-performing foundation models are much less computationally efficient than the ResNet. This problem is exacerbated by the limited digitisation of histopathology services, with most pathological diagnoses still made under a microscope. AI adoption will be contingent on it being accessible and beneficial given limited computational infrastructure and users who may not be technological experts. While various issues are inhibiting the clinical translation of ovarian cancer subtyping models, these seem increasingly likely to be overcome in the near future.



\section*{Conclusion}
In this study, we conducted a rigorous validation of feature extraction methods for ovarian cancer subtyping. We found that the features generated by histopathology foundation models drastically improved downstream classification performance when compared to ImageNet-pretrained feature extractors. 
Several different data preprocessing techniques were evaluated in an attempt to improve the performance of the ImageNet-pretrained ResNet50 baseline, and while these somewhat improved performance, they were far from sufficient to match the performance of the foundation models. Through a five-fold ensemble of ABMIL classifiers, the best overall foundation model, H-optimus-0, achieved a five-class balanced accuracy of 89\% on internal test data and 97\% and 80\% on external test sets, compared to 68\%, 81\%, and 55\% respectively for the best ImageNet-pretrained ResNet models. 
The largest models and those pretrained with the largest datasets generally gave the best performance, though the UNI foundation model was one of the best-performing models despite a relatively moderate model and dataset size.   
Hyperparameter tuning the downstream classifiers improved classification performance by a median of 1.9\% balanced accuracy, although this was variable. 
While the improved classification performance offered by histopathology foundation models may be sufficient for clinical implementation, the need to address logistical hurdles and conduct larger-scale validations remains.           

\section*{Acknowledgements}
There was no direct funding for this research. JB is supported by the UKRI Engineering and Physical Sciences Research Council (EPSRC) [EP/S024336/1]. KA is supported by the Tony Bramall Charitable Trust. The funders had no role in influencing the content of this research. There was no formal study protocol or registration, and no patient or public involvement in this research. This study was conducted retrospectively using human subject data and received approval from the Wales Research Ethics Committee [18/WA/0222] and the Confidentiality Advisory Group [18/CAG/0124]. Approval has not yet been provided for this data to be shared outside of the research group. The Transcanadian Study dataset was downloaded from \url{https://www.medicalimageanalysis.com/data/ovarian-carcinomas-histopathology-dataset} (last accessed 09/04/24). The OCEAN Challenge dataset was downloaded from \url{https://www.kaggle.com/competitions/UBC-OCEAN/data} (last accessed 20/08/24). All code used in this research is available at \url{https://github.com/scjjb/Ovarian_Features}. For the purpose of open access, the author has applied a Creative Commons Attribution (CC BY) licence to any Author Accepted Manuscript version arising from this submission. 

\section*{Author Contributions}
JB created the study protocol with feedback and contributions from all other authors. KA collected and curated the internal dataset with assistance from NMO. JB conducted all experiments with advice from NR. JB wrote the manuscript, with feedback and contributions from all other authors.

\section*{Competing Interests}
NMO's fellowship is funded by 4D Path. All other authors declare no conflicts of interest.

\newpage
\appendix
\section{Hyperparameter Tuning Details}\label{app:hyperparams}

\begin{table}[htbp]
    \centering
    \resizebox{\columnwidth}{!}{%
    \begin{tabular}{|c|cccccccccc|}
    \hline 
         \textbf{\makecell{Tuning\\Iteration}}&  \textbf{\makecell{Learning\\Rate}}&  \textbf{\makecell{Weight\\Decay}}&  \textbf{\makecell{First\\Moment\\Decay}}&  \textbf{\makecell{Second\\Moment\\Decay}}&  \textbf{\makecell{Stability\\Parameter}}&  \textbf{\makecell{LR Decay\\Patience}}&  \textbf{\makecell{LR Decay\\Factor}}&  \textbf{\makecell{Model\\Size}}& \textbf{\makecell{Drop\\Out}} &\textbf{\makecell{Max\\Patches}}\\
    \hline
        1&  \checkmark&  &  &  &  &  &  &  \checkmark&  & \\
        2&  &  &  &  &  &  &  &  &  \checkmark& \checkmark\\
        3&  &  &  \checkmark&  \checkmark&  &  &  &  &  & \\
        4&  \checkmark&  \checkmark&  &  &  &  &  &  &  & \\
        5&  &  &  \checkmark&  &  \checkmark&  &  &  &  & \\
        6&  &  &  &  &  &  &  &  \checkmark&  & \checkmark\\
        7&  &  &  &  &  &  \checkmark&  \checkmark&  &  & \\
        8&  \checkmark&  &  &  &  &  &  &  &  \checkmark& \\
        9&  &  &  &  &  &  &  &  \checkmark&  & \\
        10&  \checkmark&  &  &  &  &  \checkmark&  &  \checkmark&  & \\
        11&  &  &  &  &  &  &  &  &  \checkmark& \checkmark\\
        12&  &  &  &  &  &  \checkmark&  \checkmark&  &  & \\
        13&  \checkmark&  &  &  &  &  &  &  \checkmark&  & \\
        14&  &  \checkmark&  &  &  &  &  &  &  & \checkmark\\
        15&  &  &  &  &  &  &  &  \checkmark&  & \\
        16&  &  &  \checkmark&  \checkmark&  &  &  &  &  & \\
        17&  \checkmark&  &  \checkmark&  &  &  &  &  \checkmark&  \checkmark& \checkmark\\
    \hline 
    \end{tabular}
    }
    \caption{Iterative hyperparameter tuning procedure, with check marks (\checkmark) indicating the hyperparameters that were adjusted at each stage of tuning, with all others frozen.}
    \label{tab:tuningstages}
\end{table}

\begin{figure}[htbp]
\centering
\includegraphics[width=\textwidth]{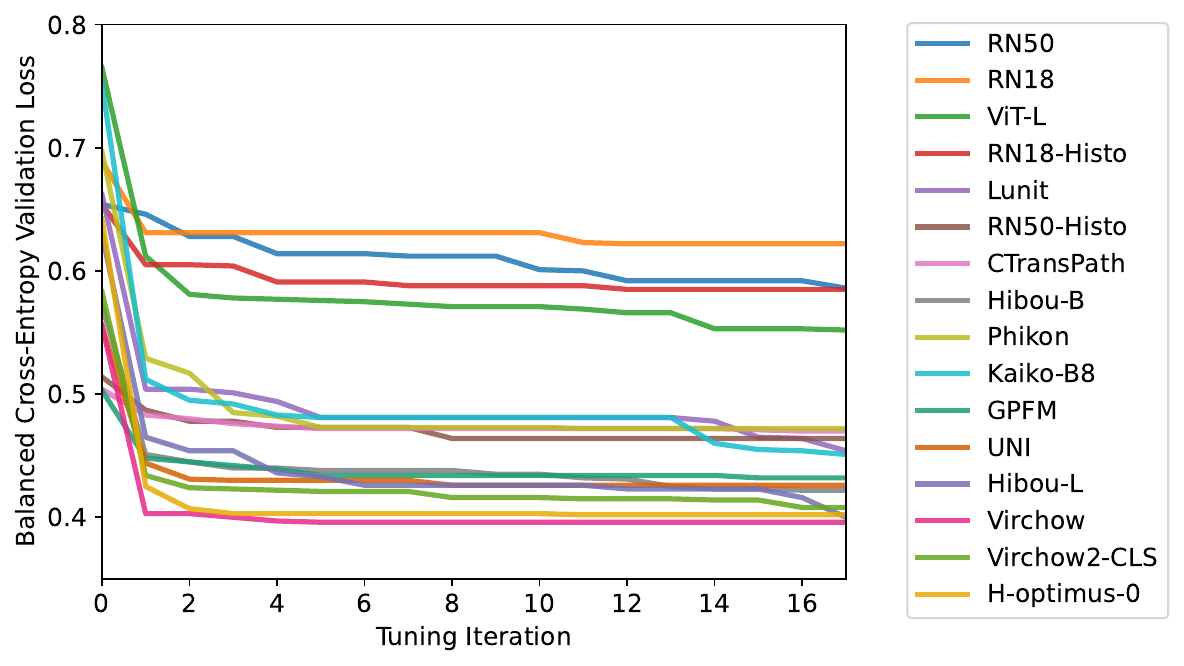}
\caption{Average validation loss from five-fold cross-validation for each model and each hyperparameter tuning iteration.}
\label{fig:tuningloss}
\end{figure}

Hyperparameter tuning was conducted over seventeen iterations for the ten ABMIL classifier hyperparameters, as shown in Table \ref{tab:tuningstages}. The optimal hyperparameters (Table \ref{tab:hyperparams}) typically did not vary greatly for models using the same feature extraction backbone, with a few notable exceptions. The regularisation hyperparameters (drop out, weight decay, max patches) varied greatly across all models, including those with the same backbone.  
The classifier based on the x5 augmented training data was the smallest ResNet50-based classifier by far (and had the smallest stability parameter and learning rate decay factor), with only 0.1M parameters compared to the next smallest at 0.7M. The ViT-based models had between 0.2M (H-optimus-0) and 1.6M parameters (Virchow). 
The largest ViT-based encoders typically had smaller values for the first moment decay (0.5-0.75) than the smaller ViT-based encoders (0.9-0.99).   Other hyperparameters were relatively stable within a given feature extractor. 

Some hyperparameters varied greatly between model architectures. The learning rate was much smaller for ViT-based models (0.00001-0.0002) than ImageNet-pretrained ResNet50 models (0.001-0.002) and often had a faster rate of decay. 
The Adam optimiser first and second moment decay parameters were also often higher in ViT-based models than in ResNet50 models. Other hyperparameters were relatively consistent between model architectures. 

As shown in Figure \ref{fig:tuningloss}, most of the benefit of hyperparameter tuning was achieved within the first few tuning iterations. This indicates the importance of tuning the key hyperparameters (especially learning rate and model size) and also indicates that hyperparameter tuning does not necessarily need to be extensive to provide a benefit. The models with larger feature extractors typically achieved smaller loss values in tuning. It is worth noting that the ABMIL classifiers were orders of magnitude smaller than the feature extraction models, making it much more computationally feasible to tune the classifiers than the feature extractors. 

\begin{table}[htbp]
    \centering
    \resizebox{\columnwidth}{!}{%
    \begin{tabular}{|c|ccc|ccc|ccc|c|}
    \hline 
         \textbf{\makecell{Feature \\ Extractor}}&  \textbf{\makecell{Learning\\Rate (LR)}}&  \textbf{\makecell{LR Decay\\Patience}}&  \textbf{\makecell{LR Decay\\Factor}}& \textbf{\makecell{First\\Moment\\Decay}}&  \textbf{\makecell{Second\\Moment\\Decay}}&  \textbf{\makecell{Stability\\Parameter}}& \textbf{\makecell{Drop\\Out}} & \textbf{\makecell{Weight\\Decay}}& \textbf{\makecell{Max\\Patches}} &    \textbf{\makecell{Model\\Size}} \\
    \hline
         RN50 &  2e-3& 20&  0.75&   0.75&  0.95&  1e-2&  0.4& 1e-3& 800&   [512,128]\\
         RN18 &  1e-4& 20&  0.9&    0.8&  0.99&  1e-4&  0.5& 1e-5& 700&    [1024,256]\\
         ViT-L& 5e-5& 10& 0.35&  0.85& 0.999& 1e-3& 0.0& 1e-1& 800 &  [512,384] \\
         \hline
         RN18-Histo& 2e-4& 20& 0.9&  0.9& 0.99& 1e-4& 0.6& 1e-4& 1000&  [512,512]\\
         Lunit & 1e-4 & 10 & 0.75 & 0.99 & 0.9999 & 1e-5 & 0.6 & 1e-1 & 900 & [1024,512] \\
         RN50-Histo & 2e-4 & 25 & 0.75 & 0.8 & 0.99 & 1e-4 & 0.6 & 1e-3 & 700 & [512,384] \\
         CTransPath & 1e-4 & 25 & 0.9 & 0.7 & 0.99999 & 1e-3 & 0.4 & 1e-3 & 1000 & [256,128] \\
         Hibou-B & 4e-5 & 10 & 0.9 & 0.99 & 0.9999 & 1e-3 & 0.3 & 1e-2 & 1600 & [256,128] \\
         Phikon & 5e-5 & 25 & 0.75 & 0.99 & 0.999 & 1e-5 & 0.8 & 1e-5 & 1200 & [512,256] \\
         Kaiko-B8 & 2e-5 & 10 & 0.75 & 0.95 & 0.9999 & 1e-5 & 0.2 & 1e-1 & 600 & [512,128] \\
          GPFM & 1e-4 & 25 & 0.9 & 0.95 & 0.99 & 1e-4 & 0.8 & 1e-6 & 1000 & [512,128] \\
          UNI & 1e-5& 10& 0.75&  0.9& 0.999& 1e-5& 0.0& 1e-3& 1000&  [512,256]\\
          Hibou-L & 5e-5 & 25 & 0.75 & 0.75 & 0.99999 & 1e-4 & 0.6 & 1e-7 & 400 & [256,128] \\
          Virchow & 2e-4 & 20 & 0.9 & 0.95 & 0.99 & 1e-3 & 0.8 & 1e-2 & 1100 & [512,256] \\
          Virchow2-CLS & 2e-5 & 10 & 0.75 & 0.55 & 0.999 & 1e-4 & 0.6 & 1e-4 & 1000 & [512,256] \\
          H-optimus-0 & 2.5e-5 & 5 & 0.75 & 0.5 & 0.9999 & 1e-4 & 0.4 & 1e-2 & 1000 & [128,32] \\
         Prov-GigaPath & 5e-5 & 15 & 0.75 & 0.7 & 0.99 & 1e-4 & 0.7 & 1e-4 & 1300 & [512,256] \\
          \hline
         RN50 Reinhard&  2e-3& 25&  0.75&    0.75&  0.95&  1e-2&  0.4& 1e-3& 400&    [512,256]\\
         RN50 Macenko&  2e-3& 15&  0.75&     0.85&  0.95&  1e-2&  0.3& 1e-3& 400&   [512,128]\\
         RN50 Otsu&  2e-3& 15&  0.9&     0.75&  0.95&  1e-2&  0.1& 1e-3& 600&   [512,256]\\
         RN50 Otsu+Macenko&  2e-3&  25&  0.9&  0.75&  0.99&  1e-3&  0.3& 1e-4&  1000&  [512,256]\\
         RN50 5Augs&  1e-3&  25&  0.6&   0.8&  0.99&  1e-4&  0.4& 1e-4&  700&    [128,32]\\
         RN50 10Augs&  2e-3&  20&  0.75&    0.8&  0.99&  1e-2&  0.4& 1e-3& 700&   [512,256]\\
         RN50 20Augs&  1e-3&  20&  0.75&   0.7&  0.999&  1e-3&  0.6& 1e-4&  1000&    [512,128]\\
         \hline 
    \end{tabular}
    }
    \caption{The final hyperparameters of each model determined by an iterative grid search tuning procedure using five cross-validation folds, including the models from the ablation study. The model size is presented as the number of parameters in the attention layer and subsequent fully connected layer. RN18 - ResNet18. RN50 - ResNet50. ViT-L - Vision Transformer.}
    \label{tab:hyperparams}
\end{table}

\clearpage
\section{Expanded Results}\label{app:results}

\begin{table}[h]
    \centering
    \resizebox{\columnwidth}{!}{%
    \begin{tabular}{|c|ccc|}
         \hline
         \textbf{\makecell{Feature \\ Extractor}}& \textbf{\makecell{Balanced\\Accuracy}}   
&  \textbf{AUROC} & \textbf{F1 Score}
\\
         \hline
         RN50& 57.1\% (53.8-60.4\%)  
& 0.893 (0.879-0.907)&  0.596 (0.561-0.630) 
\\
          RN18 & 56.1\% (52.8-59.4\%)
&  0.882 (0.866-0.898) & 0.584 (0.551-0.617)
\\
 ViT-L & 62.6\% (59.2-66.0\%)  
& 0.893 (0.877-0.909) & 0.628 (0.596-0.660)
\\
 \hline
  RN18-Histo & 59.1\% (55.8-62.4\%)
& 0.887 (0.871-0.902) &  0.615 (0.582-0.648) 
\\
Lunit & 66.6\% (63.3-70.0\%)
& 0.910 (0.894-0.926) & 0.682  (0.649-0.714) 
\\
RN50-Histo & 62.4\% (59.2-65.6\%)
& 0.925 (0.911-0.938) &  0.651 (0.618-0.684) 
\\
CTransPath & 67.3\% (63.9-70.6\%)  
& 0.925 (0.912-0.938) & 0.669 (0.638-0.700)
\\
Hibou-B & 67.7\% (64.4-71.0\%) 
& 0.945 (0.935-0.954) & 0.689 (0.656-0.720)
\\
Phikon & 67.0\% (63.7-70.4\%) 
& 0.926 (0.912-0.938) & 0.684 (0.653-0.715)
\\
Kaiko-B8 & 70.3\% (67.0-73.6\%) 
& 0.933 (0.919-0.946) & 0.720 (0.688-0.751)
\\
  GPFM & 70.9\% (67.7-74.1\%) & 0.935 (0.923-0.948) & 0.710 (0.680-0.739) \\ 
  UNI & 73.2\% (69.9-76.4\%)  & 0.945 (0.933-0.956) & 0.734 (0.704-0.764)\\
  Hibou-L & 67.0\% (63.6-70.3\%)
& 0.930 (0.918-0.942) & 0.690  (0.656-0.721) 
\\
  Virchow  & 68.6\% (65.3-71.8\%) 
& 0.936 (0.925-0.947) & 0.688 (0.658-0.717)
\\
  Virchow2-CLS  & \textbf{74.7\%} (71.5-77.9\%) 
& 0.943 (0.930-0.954) & \textbf{0.742} (0.713-0.771)
\\
H-optimus-0 & 72.2\% (68.9-75.4\%) 
& \textbf{0.947} (0.936-0.957) & 0.726 (0.695-0.756)
\\
Prov-GigaPath & 71.2\% (67.9-74.4\%) 
& 0.927 (0.913-0.941) & 0.725 (0.696-0.754)
\\
 \hline
    \end{tabular}
    }
    \caption{Results of five-fold cross-validation. Results are reported as the mean and 95\% confidence intervals (in brackets) from 10,000 iterations of bootstrapping. The greatest results are shown in \textbf{bold}.}
 \label{table:cv-results}
\end{table}

\begin{table}[h]
    \centering
    \resizebox{\columnwidth}{!}{%
    \begin{tabular}{|c|ccc|}
         \hline
         \textbf{\makecell{Feature \\ Extractor}}& \textbf{\makecell{Balanced\\Accuracy}}   
&  \textbf{AUROC} & \textbf{F1 Score}
\\
         \hline
         RN50& 66.0\% (58.1-73.7\%)  
&  0.916 (0.873-0.953)&  0.634 (0.537-0.726) 
\\
          RN18 & 64.0\% (55.3-72.6\%)
&  0.930 (0.893-0.963) &  0.628 (0.530-0.723)
\\
 ViT-L & 76.0\% (67.8-83.7\%)  
&  0.926 (0.885-0.963) &  0.747 (0.656-0.832)
\\
 \hline
  RN18-Histo & 65.0\% (57.1-72.5\%)
&  0.890 (0.843-0.932) &  0.613 (0.531-0.698) 
\\
Lunit & 79.1\% (71.4-86.3\%)
& 0.943 (0.904-0.977) & 0.778  (0.693-0.857) 
\\
RN50-Histo & 74.1\% (65.7-81.9\%)
& 0.946 (0.908-0.977) &  0.730 (0.641-0.815) 
\\
CTransPath & 81.0\% (74.0-88.0\%)  
& 0.950 (0.911-0.982) & 0.797 (0.716-0.873)
\\
Hibou-B & 87.0\% (81.0-92.6\%) 
& 0.956 (0.921-0.985) & 0.858 (0.783-0.925)
\\
Phikon & 79.0\% (72.0-85.7\%) 
& 0.946 (0.907-0.979) & 0.772 (0.689-0.852)
\\
Kaiko-B8 & 83.0\% (75.8-89.9\%)  
& 0.947 (0.909-0.980) & 0.823 (0.746-0.896)
\\
  GPFM & 82.0\% (74.8-88.7\%) & 0.955 (0.918-0.985) & 0.809 (0.728-0.884) \\ 
  UNI & 88.0\% (81.5-93.8\%)  &  0.957 (0.919-0.989) &  0.875 (0.805-0.937)\\
Hibou-L & 82.1\% (75.5-88.4\%)
& 0.959 (0.921-0.990) &  0.804 (0.722-0.880) 
\\
  Virchow &  85.0\% (78.4-91.1\%)
& \textbf{0.964} (0.928-0.993) & 0.839 (0.763-0.909)
\\
  Virchow2-CLS &  88.0\% (81.9-93.8\%)
& \textbf{0.964} (0.926-0.994) & 0.873 (0.802-0.937)
\\
H-optimus-0 & \textbf{89.0\%} (83.1-94.3\%) 
& 0.963 (0.925-0.992) & \textbf{0.883} (0.815-0.944)
\\
Prov-GigaPath & 84.0\% (77.4-90.3\%) 
& 0.958 (0.924-0.986) & 0.830 (0.752-0.900)
\\
 \hline
    \end{tabular}
    }
    \caption{Results of hold-out testing, with predictions generated by an ensemble of the five-fold classification models. Results are reported as the mean and 95\% confidence intervals (in brackets) from 10,000 iterations of bootstrapping. The greatest results are shown in \textbf{bold}.}
 \label{table:holdout-results}
\end{table}

\begin{table}[h]
    \centering
    \resizebox{\columnwidth}{!}{%
    \begin{tabular}{|c|ccc|}
         \hline
         \textbf{\makecell{Feature \\ Extractor}}& \textbf{\makecell{Balanced\\Accuracy}}   
&  \textbf{AUROC} & \textbf{F1 Score}
\\
         \hline
         RN50& 69.2\% (58.7-79.7\%)  
&  0.956 (0.928-0.980)&   0.696 (0.582-0.807) 
\\
          RN18 & 79.0\% (68.8-88.6\%)
&   0.959 (0.923-0.985) &  0.804 (0.700-0.896)
\\
 ViT-L & 80.7\% (72.2-89.2\%)  
&  0.970 (0.937-0.993) &  0.814 (0.712-0.908)
\\
 \hline
  RN18-Histo & 66.5\% (55.2-77.5\%)
&  0.930 (0.888-0.965) &   0.653 (0.539-0.763) 
\\
Lunit & 95.0\% (89.3-99.1\%)
& 0.998 (0.994-1.000) & 0.930  (0.862-0.985) 
\\
RN50-Histo & 94.4\% (88.2-98.9\%)
& 0.994 (0.985-0.999) &  0.934 (0.870-0.985) 
\\
CTransPath & 88.8\% (80.9-95.6\%)  
& 0.982 (0.959-0.996) & 0.861 (0.773-0.939)
\\
Hibou-B & 91.1\% (83.0-97.9\%) 
& 0.990 (0.979-0.998) & 0.921 (0.850-0.979)
\\
Phikon & 90.3\% (81.9-97.8\%) 
& 0.994 (0.986-0.999) & 0.919 (0.839-0.982)
\\
Kaiko-B8 & 96.7\% (93.8-99.2\%) 
& 0.997 (0.991-1.000) & 0.937 (0.879-0.986)
\\
  GPFM & \textbf{98.3\%} (95.6-100.0\%) & \textbf{0.999} (0.997-1.000) & \textbf{0.977} (0.937-1.000) \\ 
  UNI & 93.2\% (86.5-98.3\%)  &  0.996 (0.988-1.000) &  0.912 (0.835-0.974) \\
  Hibou-L & 89.3\% (80.6-96.2\%)
& 0.989 (0.975-0.998) &  0.889 (0.805-0.959) 
\\
  Virchow & 87.5\% (79.0-94.8\%) 
& 0.993 (0.984-0.999) & 0.848 (0.750-0.931)
\\
  Virchow2-CLS & 88.0\% (79.8-95.3\%) 
& 0.997 (0.993-1.000) & 0.871 (0.779-0.952)
\\
H-optimus-0 & 96.7\% (91.1-100.0\%) 
& \textbf{0.999} (0.998-1.000) & 0.975 (0.931-1.000)
\\
Prov-GigaPath & 88.6\% (80.2-95.9\%) 
& 0.995 (0.987-1.000) & 0.878 (0.783-0.958)
\\
 \hline
    \end{tabular}
    }
    \caption{Results of external validation on the Transcanadian Study dataset, with predictions generated by an ensemble of the five-fold classification models. Results are reported as the mean and 95\% confidence intervals (in brackets) from 10,000 iterations of bootstrapping. The greatest results are shown in \textbf{bold}.}
 \label{table:external-results}
\end{table}

\begin{table}[h]
    \centering
    \resizebox{\columnwidth}{!}{%
    \begin{tabular}{|c|ccc|}
         \hline
         \textbf{\makecell{Feature \\ Extractor}}& \textbf{\makecell{Balanced\\Accuracy}}   
&  \textbf{AUROC} & \textbf{F1 Score}
\\
         \hline
  RN50 & 52.4\% (49.5-55.1\%)  &  0.868 (0.847-0.889) &  0.412 (0.380-0.444) \\
  RN18 & 51.9\% (48.6-54.9\%)  & 0.841  (0.820-0.863) &  0.412 (0.377-0.448) \\
  ViT-L & 59.5\% (55.4-63.6\%)  & 0.880  (0.857-0.902) &  0.578 (0.532-0.625) \\
  \hline
  RN18-Histo & 57.3\% (54.0-60.4\%)  &  0.850 (0.828-0.872) &  0.523 (0.484-0.563) \\ 
  Lunit & 73.6\% (69.7-77.5\%)
& 0.954 (0.941-0.967) &  0.729 (0.681-0.775) 
\\
RN50-Histo & 68.0\% (64.5-71.6\%)
& 0.946 (0.930-0.959) & 0.679  (0.634-0.725) 
\\
  CTransPath & 67.8\% (64.0-71.7\%)  &  0.934 (0.917-0.950) & 0.676  (0.629-0.724) \\
  Hibou-B & 65.4\% (61.4-69.4\%)  & 0.935  (0.920-0.949) &  0.633 (0.582-0.682) \\
  Phikon & 66.4\% (62.8-70.1\%)  &  0.898 (0.879-0.917) & 0.642  (0.595-0.689) \\
  Kaiko-B8 & 70.0\% (65.4-74.5\%)  &  0.941 (0.925-0.956) &  0.695 (0.644-0.744) \\
  GPFM & 74.5\% (70.4-78.5\%) & 0.935 (0.919-0.949) & 0.746 (0.702-0.788) \\ 
  UNI & 77.2\% (73.0-81.4\%)  &  0.954 (0.939-0.966) &  0.758 (0.714-0.801) \\
    Hibou-L & 69.3\% (66.2-72.3\%)
& 0.946 (0.931-0.959) &  0.663 (0.622-0.706) 
\\
  Virchow & 79.2\% (75.2-83.0\%)  &  \textbf{0.959} (0.946-0.970) & \textbf{0.765}  (0.722-0.807) \\
  Virchow2-CLS & \textbf{79.8\%} (75.8-83.6\%)  &  0.958 (0.945-0.970) &  0.759 (0.717-0.801) \\
  H-optimus-0 & 74.0\% (69.9-78.1\%)  &  0.952 (0.939-0.963) &  0.703 (0.656-0.748) \\
  Prov-GigaPath & 75.4\% (71.3-79.3\%)  & \textbf{0.959}  (0.946-0.970) &  0.729 (0.684-0.771) \\
 \hline
    \end{tabular}
    }
    \caption{Results of external validation on the OCEAN dataset, with predictions generated by an ensemble of the five-fold classification models. Results are reported as the mean and 95\% confidence intervals (in brackets) from 10,000 iterations of bootstrapping. The greatest results are shown in \textbf{bold}.}
 \label{table:ocean-results}
\end{table}

\clearpage

\section{Hyperparameter Tuning Ablation}\label{app:tuningablation}

The results of each model using the default hyperparameters (without tuning) are shown in Tables \ref{table:untunedCV}-\ref{table:untunedOcean}, with indications about whether the default hyperparameters improved or degraded performance compared to using the tuned hyperparameters. Exact p-values are provided in \ref{app:stats}. 

\begin{table}[h]
    \centering
    \begin{tabular}{|c|cc|cc|cc|}
         \hline
         \textbf{\makecell{Feature \\ Extractor}}& \multicolumn{2}{c|}{\textbf{\makecell{Balanced\\Accuracy}}}  
&  \multicolumn{2}{c|}{\textbf{AUROC}} & \multicolumn{2}{c|}{\textbf{F1 Score}} 
\\
         \hline
         RN50&   
 57.5\% & - \hspace{0.15em}
& 0.877 & $\downarrow$ \hspace{0.15em} &  0.593 & - \hspace{0.15em}   
\\
          RN18 &  55.3\% & - \hspace{0.15em} &  0.857 & $\downarrow$ \hspace{0.15em} & 0.561 & $\downarrow$ \hspace{0.15em}
\\
 ViT-L & 56.2\% & $\downarrow\downarrow\downarrow$*   
& 0.857 & $\downarrow\downarrow$*  & 0.580 &$\downarrow\downarrow$* 
\\
 \hline
  RN18-Histo & 56.0\% & $\downarrow\downarrow$* 
& 0.879 & - \hspace{0.15em}  &  0.574 & $\downarrow\downarrow$*  
\\
Lunit & 65.3\% & $\downarrow$ \hspace{0.15em} 
& 0.891 &  $\downarrow$*  & 0.646 & $\downarrow\downarrow$ \hspace{0.15em} 
\\
RN50-Histo & 63.1\% & - \hspace{0.15em} 
& 0.915 & $\downarrow$ \hspace{0.15em}  & 0.656 & - \hspace{0.15em} 
\\
CTransPath & 68.8\% & $\uparrow$  \hspace{0.15em} 
& 0.927 & - \hspace{0.15em} & 0.690 & $\uparrow$ \hspace{0.15em}
\\
Hibou-B & 66.1\%  & $\downarrow$ \hspace{0.15em} 
& 0.911 & $\downarrow\downarrow$ \hspace{0.15em} & 0.667 & $\downarrow$ \hspace{0.15em}
\\
Phikon & 68.0\% & $\uparrow$ \hspace{0.15em}  
& 0.912 & $\downarrow$ \hspace{0.15em}   & 0.672 & $\downarrow$ \hspace{0.15em}
\\
Kaiko-B8 & 62.7\%  
& $\downarrow\downarrow\downarrow$* & 0.907 & $\downarrow$ \hspace{0.15em} & 0.633 & $\downarrow\downarrow\downarrow$*
\\
  GPFM & 69.4\% & $\downarrow$ \hspace{0.15em} & 0.912 & $\downarrow$* & 0.690 & $\downarrow$ \hspace{0.15em} \\
  UNI & 67.1\%   &  $\downarrow\downarrow\downarrow$* & 0.915 & $\downarrow\downarrow$*  & 0.684  &$\downarrow\downarrow\downarrow$* \\
  Hibou-L & 58.7\% & $\downarrow\downarrow\downarrow$ \hspace{0.15em} 
& 0.889 &  $\downarrow\downarrow$*  & 0.622 & $\downarrow\downarrow\downarrow$ \hspace{0.15em} 
\\
  Virchow & 65.2\%  
& $\downarrow\downarrow$* & 0.896 & $\downarrow\downarrow$ \hspace{0.15em} & 0.658 & $\downarrow\downarrow$ \hspace{0.15em} \\
  Virchow2-CLS & 69.8\% & $\downarrow\downarrow$ \hspace{0.15em} & 0.917 & $\downarrow$* & 0.681 & $\downarrow\downarrow\downarrow$ \hspace{0.15em}
\\
H-optimus-0 & 66.1\%  
& $\downarrow\downarrow\downarrow$ \hspace{0.15em} & 0.916 & $\downarrow\downarrow$ \hspace{0.15em} & 0.678 & $\downarrow\downarrow$ \hspace{0.15em}
\\
Prov-GigaPath & 67.9\%  
& $\downarrow\downarrow$ \hspace{0.15em} & 0.919 & - \hspace{0.15em} & 0.675 & $\downarrow\downarrow\downarrow$ \hspace{0.15em}
\\
 \hline
    \end{tabular}
    \caption{Results of five-fold cross-validation without hyperparameter tuning. Arrows indicate the difference in performance relative to the tuned models, with one arrow ($\uparrow$) for difference a of at least 1\%, two arrows ($\uparrow\uparrow$) for a difference of at least 3\%, and three arrows ($\uparrow\uparrow\uparrow$) for a difference of at least 5\%. *Indicates a p-value less than 0.05 when comparing the cross-validation folds to the tuned model.}
 \label{table:untunedCV}
\end{table}

\begin{table}[h]
    \centering
    \begin{tabular}{|c|cc|cc|cc|}
         \hline
         \textbf{\makecell{Feature \\ Extractor}}& \multicolumn{2}{c|}{\textbf{\makecell{Balanced\\Accuracy}}}  
&  \multicolumn{2}{c|}{\textbf{AUROC}} & \multicolumn{2}{c|}{\textbf{F1 Score}} 
\\
         \hline
         RN50&  68.0\%  
  & $\uparrow$ \hspace{0.15em}
& 0.923 & - \hspace{0.15em} & 0.670  & $\uparrow\uparrow$ \hspace{0.15em}   
\\
          RN18 & 64.0\%  & - \hspace{0.15em} 
&  0.927 &  - \hspace{0.15em} & 0.626 & - \hspace{0.15em}
\\
 ViT-L & 61.0\% & $\downarrow\downarrow\downarrow$*     
& 0.917 & - \hspace{0.15em} & 0.601 & $\downarrow\downarrow\downarrow$* 
\\
 \hline
  RN18-Histo & 62.1\% & $\downarrow$ \hspace{0.15em}   
& 0.889 & - \hspace{0.15em} & 0.586  & $\downarrow$  \hspace{0.15em}
\\
Lunit & 74.0\% & $\downarrow\downarrow\downarrow$ \hspace{0.15em} 
& 0.932 & $\downarrow$ \hspace{0.15em}  & 0.727  & $\downarrow\downarrow\downarrow$ \hspace{0.15em}  
\\
RN50-Histo & 74.9\% & - \hspace{0.15em}
& 0.943 & - \hspace{0.15em}   & 0.739 & - \hspace{0.15em}  
\\
CTransPath & 78.0\% & $\downarrow\downarrow$*    
& 0.941 &  - \hspace{0.15em} & 0.768 & $\downarrow$* 
\\
Hibou-B &  78.0\% & $\downarrow\downarrow\downarrow$*
& 0.958 & - \hspace{0.15em} & 0.765 & $\downarrow\downarrow\downarrow$* 
\\
Phikon & 80.1\% &  $\uparrow$ \hspace{0.15em} 
& 0.941  &  - \hspace{0.15em}  & 0.792 & $\uparrow$ \hspace{0.15em}
\\
Kaiko-B8 &  79.0\% 
& $\downarrow\downarrow$ \hspace{0.15em} & 0.949 & - \hspace{0.15em} & 0.786 & $\downarrow\downarrow$ \hspace{0.15em}
\\
  GPFM & 84.0\% & $\uparrow$ \hspace{0.15em} & 0.953 & - \hspace{0.15em} & 0.831 & $\uparrow$ \hspace{0.15em} \\
  UNI & 82.0\%   &   $\downarrow\downarrow\downarrow$*  & 0.962 & - \hspace{0.15em} & 0.806 & $\downarrow\downarrow\downarrow$*  \\
  Hibou-L & 75.0\% & $\downarrow\downarrow\downarrow$* 
& 0.959 & - \hspace{0.15em}  & 0.730 & $\downarrow\downarrow\downarrow$* 
\\
  Virchow & 81.0\%  
& $\downarrow\downarrow$ \hspace{0.15em}  & 0.956 & - \hspace{0.15em} & 0.801 & $\downarrow\downarrow$ \hspace{0.15em}
\\
  Virchow2-CLS & 89.1\% & $\uparrow$ \hspace{0.15em} & 0.963 & - \hspace{0.15em} & 0.883 & $\uparrow$ \hspace{0.15em}
\\
H-optimus-0 & 85.0\%  
& $\downarrow\downarrow$ \hspace{0.15em}  & 0.965 & - \hspace{0.15em} & 0.843 & $\downarrow\downarrow$ \hspace{0.15em}
\\
Prov-GigaPath & 83.0\%  
&  $\downarrow$ \hspace{0.15em} & 0.949 & -* & 0.820 & $\downarrow$ \hspace{0.15em}
\\
 \hline
    \end{tabular}
    \caption{Results of hold-out testing without hyperparameter tuning. Arrows indicate the difference in performance relative to the tuned models, with one arrow ($\uparrow$) for difference a of at least 1\%, two arrows ($\uparrow\uparrow$) for a difference of at least 3\%, and three arrows ($\uparrow\uparrow\uparrow$) for a difference of at least 5\%. *Indicates a p-value less than 0.05 when comparing the cross-validation folds to the tuned model. While the Prov-GigaPath AUROC only exhibited a reduction of 0.009, this was found to be statistically significant.}
 \label{table:untunedHoldout}
\end{table}

\begin{table}[h]
    \centering
    \begin{tabular}{|c|cc|cc|cc|}
         \hline
         \textbf{\makecell{Feature \\ Extractor}}& \multicolumn{2}{c|}{\textbf{\makecell{Balanced\\Accuracy}}}  
&  \multicolumn{2}{c|}{\textbf{AUROC}} & \multicolumn{2}{c|}{\textbf{F1 Score}} 
\\
         \hline
         RN50&  75.8\% & $\uparrow\uparrow\uparrow$ \hspace{0.15em}  
   
& 0.969 & $\uparrow\uparrow$ \hspace{0.15em} & 0.769  & $\uparrow\uparrow\uparrow$  \hspace{0.15em} 
\\
          RN18 & 80.1\%  & $\uparrow$ \hspace{0.15em}
& 0.946  &  $\downarrow$ \hspace{0.15em} & 0.807 & - \hspace{0.15em}
\\
 ViT-L & 68.9\% &  $\downarrow\downarrow\downarrow$*   
& 0.960 & $\downarrow$ \hspace{0.15em} & 0.702 & $\downarrow\downarrow\downarrow$*
\\
 \hline
  RN18-Histo & 69.3\% &  $\uparrow$ \hspace{0.15em}
& 0.942 & $\uparrow$ \hspace{0.15em} & 0.689  & $\uparrow\uparrow$  \hspace{0.15em}
\\
Lunit & 92.4\% & $\downarrow$ \hspace{0.15em} 
& 0.989 & - \hspace{0.15em}  & 0.879 &  $\downarrow\downarrow\downarrow$ \hspace{0.15em} 
\\
RN50-Histo & 91.8\% & $\downarrow$ \hspace{0.15em}
& 0.995 & - \hspace{0.15em}  & 0.902 & $\downarrow\downarrow$ \hspace{0.15em} 
\\
CTransPath & 88.1\% &  -  \hspace{0.15em}
& 0.978 &  - \hspace{0.15em} & 0.847 & $\downarrow$ \hspace{0.15em}
\\
Hibou-B &  85.3\% & $\downarrow\downarrow\downarrow$ \hspace{0.15em}
& 0.987 & - \hspace{0.15em} & 0.871 & $\downarrow\downarrow\downarrow$ \hspace{0.15em}
\\
Phikon & 94.6\% &  $\uparrow\uparrow$ \hspace{0.15em} 
& 0.996 &  - \hspace{0.15em} & 0.944 & $\uparrow$ \hspace{0.15em}
\\
Kaiko-B8 &  93.3\% 
& $\downarrow\downarrow$ \hspace{0.15em} & 0.996 & - \hspace{0.15em} & 0.926 & $\downarrow$ \hspace{0.15em}
\\
  GPFM & 97.7\% & - \hspace{0.15em} & 0.998 & - \hspace{0.15em} & 0.964 & $\downarrow$ \hspace{0.15em} \\
  UNI & 95.3\%   & $\uparrow$ \hspace{0.15em} & 0.996 & - \hspace{0.15em} & 0.952 & $\uparrow\uparrow$ \hspace{0.15em} \\
  Hibou-L & 81.7\% & $\downarrow\downarrow\downarrow$ \hspace{0.15em} 
& 0.988 & -*  & 0.813 & $\downarrow\downarrow\downarrow$*  
\\
  Virchow & 88.8\%  
& $\uparrow$ \hspace{0.15em}  & 0.991 & - \hspace{0.15em} & 0.864 & $\uparrow$ \hspace{0.15em}
\\
Virchow2-CLS & 91.6\% & $\uparrow\uparrow$ \hspace{0.15em} & 1.000 & - \hspace{0.15em} & 0.915 & $\uparrow\uparrow$ \hspace{0.15em}
\\
H-optimus-0 & 99.0\%  
&  $\uparrow$ \hspace{0.15em} & 1.000 & - \hspace{0.15em} & 0.991 & $\uparrow$ \hspace{0.15em}
\\
Prov-GigaPath & 89.4\%  
&  - \hspace{0.15em} & 0.993 & - \hspace{0.15em} & 0.871 & - \hspace{0.15em}
\\
 \hline
    \end{tabular}
    \caption{Results of validation on the Transcanadian Study dataset without hyperparameter tuning. Arrows indicate the difference in performance relative to the tuned models, with one arrow ($\uparrow$) for difference a of at least 1\%, two arrows ($\uparrow\uparrow$) for a difference of at least 3\%, and three arrows ($\uparrow\uparrow\uparrow$) for a difference of at least 5\%. *Indicates a p-value less than 0.05 when comparing the cross-validation folds to the tuned model.}
 \label{table:untunedExternal}
\end{table}

\begin{table}[h]
    \centering
    \begin{tabular}{|c|cc|cc|cc|}
         \hline
         \textbf{\makecell{Feature \\ Extractor}}& \multicolumn{2}{c|}{\textbf{\makecell{Balanced\\Accuracy}}}  
&  \multicolumn{2}{c|}{\textbf{AUROC}} & \multicolumn{2}{c|}{\textbf{F1 Score}} 
\\
         \hline
RN50 & 51.0\%  
&  $\downarrow$ \hspace{0.15em} & 0.857 & $\downarrow$ \hspace{0.15em} & 0.411 & - \hspace{0.15em}
\\
RN18 & 49.0\%  
&  $\downarrow$ \hspace{0.15em} & 0.837 & - \hspace{0.15em} & 0.370 & $\downarrow\downarrow$ \hspace{0.15em}
\\
ViT-L & 48.3\%  
&  $\downarrow\downarrow\downarrow$* & 0.843 & $\downarrow\downarrow$* & 0.506 & $\downarrow\downarrow\downarrow$*
\\
\hline
RN18-Histo & 55.0\%  
&  $\downarrow$ \hspace{0.15em} & 0.849 & - \hspace{0.15em} & 0.504 & $\downarrow$ \hspace{0.15em}
\\
Lunit & 68.8\% & $\downarrow\downarrow$ \hspace{0.15em}
& 0.935 & $\downarrow$*   & 0.661 &  $\downarrow\downarrow\downarrow$ \hspace{0.15em}
\\
RN50-Histo & 66.6\% & $\downarrow$ \hspace{0.15em} 
& 0.934 &  $\downarrow$*   & 0.684 & -   \hspace{0.15em}
\\
CTransPath & 68.1\%  
& - \hspace{0.15em}  & 0.934 & - \hspace{0.15em} & 0.686 & $\uparrow$ \hspace{0.15em}
\\
Hibou-B & 64.0\%  
& $\downarrow$ \hspace{0.15em}  & 0.928 & - \hspace{0.15em} & 0.604 & $\downarrow$ \hspace{0.15em}
\\
Phikon & 63.2\%  
&  $\downarrow\downarrow$ \hspace{0.15em} & 0.903 & - \hspace{0.15em} & 0.598 & $\downarrow\downarrow$ \hspace{0.15em}
\\
Kaiko-B8 & 64.1\%  
& $\downarrow\downarrow\downarrow$ \hspace{0.15em}  & 0.929 & $\downarrow$* & 0.596 & $\downarrow\downarrow$*
\\
GPFM & 73.8\% & - \hspace{0.15em} & 0.937 & - \hspace{0.15em} & 0.725 & $\downarrow$ \hspace{0.15em} \\
UNI & 69.6\%  
&  $\downarrow\downarrow\downarrow$* & 0.948 & -* & 0.693 & $\downarrow\downarrow\downarrow$*
\\
Hibou-L & 64.7\% & $\downarrow\downarrow$ \hspace{0.15em} 
& 0.936 & $\downarrow$*   & 0.615 & $\downarrow\downarrow$ \hspace{0.15em}  
\\
Virchow & 78.5\%  
& -*  & 0.948 & $\downarrow$* & 0.766 & - \hspace{0.15em}
\\
Virchow2-CLS & 74.0\% & $\downarrow\downarrow\downarrow$* & 0.956 & - \hspace{0.15em} & 0.719 & $\downarrow\downarrow$*
\\
H-optimus-0 & 74.8\%  
&  - \hspace{0.15em} &  0.951  & - \hspace{0.15em} & 0.747 & $\uparrow\uparrow$ \hspace{0.15em}
\\
Prov-GigaPath & 75.4\%  
&  - \hspace{0.15em} & 0.957 & - \hspace{0.15em} & 0.720 & - \hspace{0.15em}
\\
 \hline
    \end{tabular}
    \caption{Results of validation on the OCEAN Challenge dataset without hyperparameter tuning. Arrows indicate the difference in performance relative to the tuned models, with one arrow ($\uparrow$) for difference a of at least 1\%, two arrows ($\uparrow\uparrow$) for a difference of at least 3\%, and three arrows ($\uparrow\uparrow\uparrow$) for a difference of at least 5\%. *Indicates a p-value less than 0.05 when comparing the cross-validation folds to the tuned model. The UNI AUROC only exhibited a reduction of 0.006, but this was found to be statistically significant. The Virchow balanced accuracy only exhibited a reduction of 0.7\%, but this was found to be statistically significant. }
 \label{table:untunedOcean}
\end{table}

\clearpage
\section{Augmentation and Normalisation Analysis}\label{app:augmentation}

\subsection*{Preprocessing Techniques}
    
Preprocessing techniques were compared using the baseline ImageNet-pretrained ResNet50 encoder 
with the default feature extraction settings from the CLAM repository \cite{Lu2021} (static saturation thresholding with no augmentation or normalisation). 
Comparisons were made using Reinhard normalisation, Macenko normalisation, Otsu thresholding, Otsu thresholding with Macenko normalisation, and colour augmentations to increase the effective training set size by factors of 5x, 10x, and 20x.


Otsu thresholding \cite{Otsu1975} is applied during tissue detection to automatically determine the saturation threshold for each image by minimising the variance within the separated high-saturation and low-saturation groups. Saturation thresholding is a computationally efficient tissue segmentation approach, but risks including artifacts such as bubbles, pen marks, and coverslip edges in the foreground. While more robust (and complex) tissue segmentation techniques exist \cite{Janowczyk2019,Shakhawat2023}, we focused on simple approaches as the attention mechanism in the classification models should learn to ignore any remaining artifacts. We compared the CLAM \cite{Lu2021} default static threshold (8/255) to Otsu thresholding with parameters manually adjusted to qualitatively improve the segmentation. 

Normalisation and augmentation techniques control data variability, which is particularly important for generalisability in histopathology, where varied staining and scanning procedures between labs result in chromatic differences \cite{Breen2023}. Normalisation reduces variability, adjusting images into a consistent colour space to allow models to learn general features. We investigated two commonly used \cite{Breen2024} stain normalisation techniques - Reinhard normalisation \cite{Reinhard2001} and Macenko normalisation \cite{Macenko2009}. These approaches work in logarithmic colour spaces, where stains behave linearly, making them easier to separate and manipulate. Reinhard normalisation is a standard normalisation technique applied in $l\alpha\beta$ space (radiance $l$, blue-yellow $\alpha$, red-green $\beta$). 
Macenko normalisation uses singular value decomposition to separate stain and saturation values, before scaling stain values in logarithmic RGB space. Basic RGB normalisations were also applied to all images (after any other colour adjustments) to match the ImageNet and histopathology-specific pretraining procedures. While many more sophisticated stain normalisation techniques have been developed, it is unclear whether any such approach is better than Macenko normalisation overall \cite{Breen2024}.  

\begin{figure}[htbp]
\centering
\includegraphics[width=\textwidth]{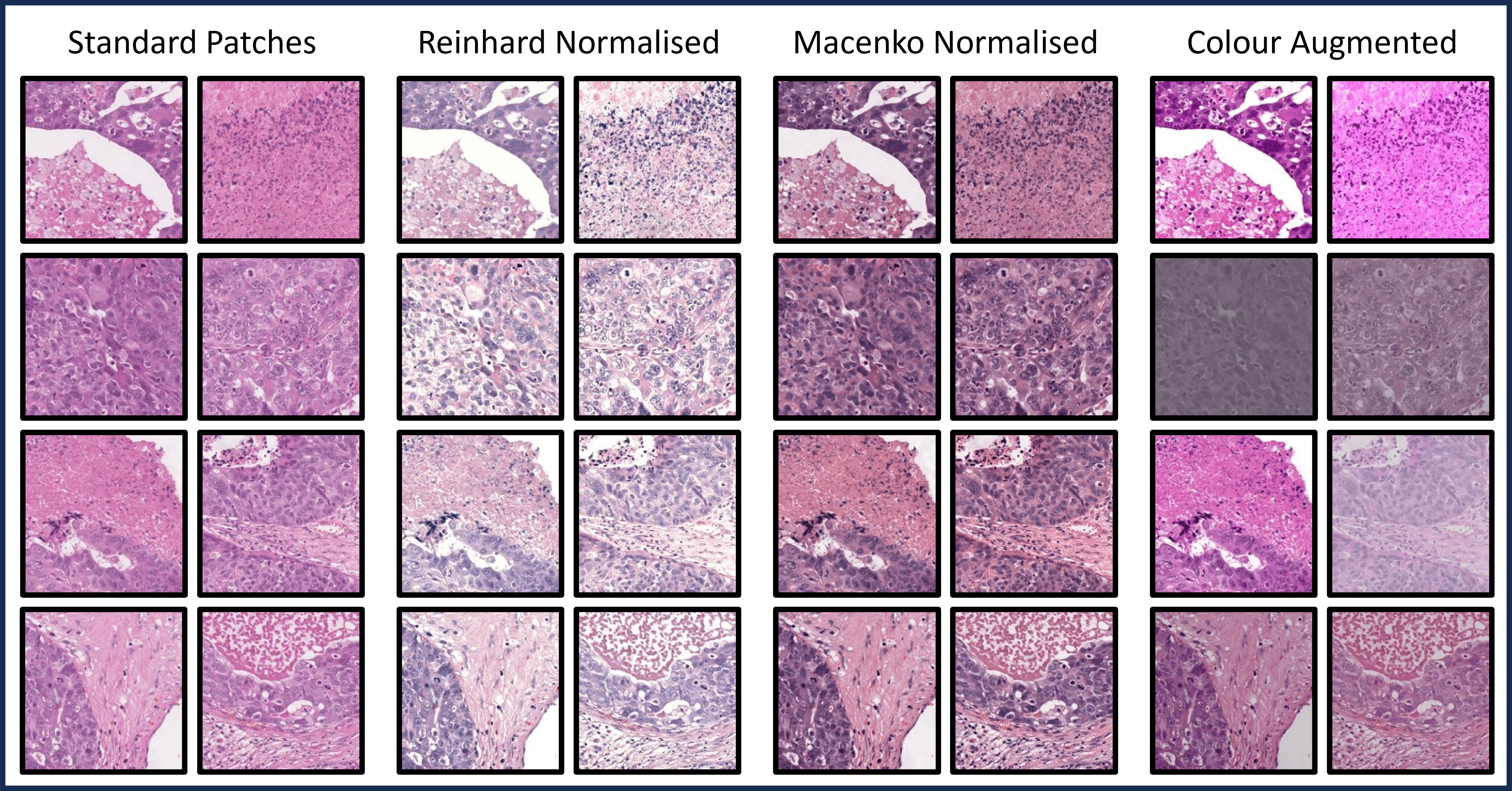}
\caption{Tissue normalisation and augmentation procedures illustrated using 256x256 pixel patches from a single whole slide image at 10x magnification.}
\label{fig:patchnorm}
\end{figure}

Augmentation techniques conversely increase the variability of the training data to allow the model to learn a more general domain. 
For such large images, training models end-to-end to allow for online data augmentation (adjustments during training) is extremely computationally intensive \cite{Dooper2023}. 
Some researchers have attempted to apply online augmentations in the embedding space using generative models \cite{Shao2023,Zaffar2023}, though this adds an extra layer of complexity to an already resource-intensive model pipeline. 
Instead, offline augmentation creates a finite set of augmented versions of the original data, artificially increasing the diversity of training data to a lesser extent than online augmentation. We investigated colour augmentations which adjusted the brightness, contrast, saturation and hue of each patch using parameters from a previous study \cite{Wang2020}, which we found to create plausibly altered colours (Figure \ref{fig:patchnorm}). 
    
\subsection*{Extended Results and Discussion}

\begin{table}[h]
    \centering
    \resizebox{\columnwidth}{!}{%
    \begin{tabular}{|c|ccc|}
         \hline
         \textbf{\makecell{Preprocessing \\ Approach}}&  \textbf{\makecell{Balanced\\Accuracy}} 
&  \textbf{AUROC}&\textbf{F1 Score}
\\
         \hline
          Baseline& 57.1\% (53.8-60.4\%)  
& \textbf{0.893} (0.879-0.907)& 0.596 (0.561-0.630)
\\
          Reinhard Normalisation& 51.3\% (48.2-54.4\%) 
&  0.872 (0.856-0.887)&  0.520 (0.488-0.553)
\\
          Macenko Normalisation& 57.8\% (54.5-61.2\%) 
& 0.882 (0.867-0.896) &  0.601 (0.567-0.635)
\\
          Otsu Thresholding& 53.9\% (50.6-57.2\%) 
& 0.888 (0.873-0.903) &   0.566 (0.532-0.600)
\\
          Otsu + Macenko& 58.0\% (54.6-61.4\%) 
& 0.882 (0.865-0.898) &  0.605 (0.571-0.638)
\\
          5x Colour Augmentation& 57.4\% (54.0-60.7\%) 
&  0.888 (0.873-0.902)&   0.592 (0.560-0.625)
\\
          10x Colour Augmentation& \textbf{59.1\%} (55.7-62.4\%) 
& 0.891 (0.877-0.905) & \textbf{0.615} (0.581-0.649)
\\
          20x Colour Augmentation& \textbf{59.1\%} (55.7-62.4\%)  
& 0.892 (0.877-0.905) & 0.596  (0.564-0.627)
\\
         \hline
    \end{tabular}
    }
    \caption{Results of five-fold cross-validation for the ImageNet-pretrained ResNet50 with varied preprocessing approaches. Results are reported as the mean and 95\% confidence intervals (in brackets) from 10,000 iterations of bootstrapping. The greatest results are shown in \textbf{bold}.}
 \label{table:cv-results-preprocessing}
\end{table}

\begin{table}[h]
    \centering
    \resizebox{\columnwidth}{!}{%
    \begin{tabular}{|c|ccc|}
         \hline
         \textbf{\makecell{Preprocessing \\ Approach}}&  \textbf{\makecell{Balanced\\Accuracy}} 
&  \textbf{AUROC}&\textbf{F1 Score}
\\
         \hline
          Baseline& 66.0\% (58.1-73.7\%)   
& 0.916 (0.873-0.953)& 0.634 (0.537-0.726)
\\
          Reinhard Normalisation& 65.0\% (56.6-73.2\%)  
& \textbf{0.923} (0.881-0.961)& 0.632 (0.534-0.727)
\\
          Macenko Normalisation& 63.0\% (54.4-71.5\%)  
& 0.915 (0.873-0.951) &  0.620 (0.521-0.715)
\\
          Otsu Thresholding& 65.0\% (56.7-73.4\%)  
& 0.916 (0.872-0.955) & 0.637 (0.542-0.732)
\\
          Otsu + Macenko& 59.0\% (50.3-67.6\%)  
& 0.918 (0.878-0.952) & 0.577 (0.475-0.674)
\\
          5x Colour Augmentation& 65.0\% (57.0-72.9\%)  
&  0.916 (0.876-0.951)& 0.630 (0.536-0.725)
\\
          10x Colour Augmentation& 64.0\% (55.9-72.1\%)   
& 0.906 (0.864-0.944) & 0.616 (0.522-0.710)
\\
          20x Colour Augmentation& \textbf{68.0\%} (59.7-76.0\%)  
& 0.904 (0.861-0.942) & \textbf{0.657} (0.563-0.750)
\\ 
         \hline
    \end{tabular}
    }
    \caption{Results of hold-out testing for the ImageNet-pretrained ResNet50 with varied preprocessing approaches, with predictions generated by an ensemble of the five-fold classification models. Results are reported as the mean and 95\% confidence intervals (in brackets) from 10,000 iterations of bootstrapping. The greatest results are shown in \textbf{bold}.}
    \label{table:holdout-results-preprocessing}
\end{table}

\begin{table}[h]
    \centering
    \resizebox{\columnwidth}{!}{%
    \begin{tabular}{|c|ccc|}
         \hline
         \textbf{\makecell{Preprocessing \\ Approach}}&  \textbf{\makecell{Balanced\\Accuracy}} 
&  \textbf{AUROC}&\textbf{F1 Score}
\\
         \hline
          Baseline& 69.2\% (58.7-79.7\%)   
& 0.956 (0.928-0.980) & 0.696 (0.582-0.807)
\\
           Reinhard Normalisation& 75.8\% (65.1-86.0\%)   
& 0.968 (0.943-0.986) & 0.761 (0.647-0.861)
\\
           Macenko Normalisation& 74.5\% (64.3-84.3\%)  
& 0.959 (0.933-0.980) & 0.756 (0.648-0.857)
\\
           Otsu Thresholding& 77.2\% (66.4-87.6\%) 
& 0.963 (0.937-0.985) &  0.797 (0.685-0.895)
\\
           Otsu + Macenko& \textbf{80.5\%} (70.4-89.9\%)  
& \textbf{0.983} (0.967-0.995) & \textbf{0.834} (0.730-0.921)
\\
           5x Colour Augmentation& 74.9\% (63.8-85.6\%)   
& 0.966 (0.941-0.986) & 0.762 (0.647-0.866)
\\
           10x Colour Augmentation& 76.1\% (65.0-86.6\%)  
& 0.962 (0.935-0.983) & 0.768 (0.659-0.869)
\\
           20x Colour Augmentation& 80.0\% (69.2-90.0\%)  
& 0.973 (0.953-0.989) & 0.806 (0.706-0.897)
\\
         \hline
    \end{tabular}
    }
    \caption{Results of external validation on the Transcanadian Study dataset for the ImageNet-pretrained ResNet50 with varied preprocessing approaches, with predictions generated by an ensemble of the five-fold classification models. Results are reported as the mean and 95\% confidence intervals (in brackets) from 10,000 iterations of bootstrapping. The greatest results are shown in \textbf{bold}.}
    \label{table:external-results-preprocessing}
\end{table}

\begin{table}[h]
    \centering
    \resizebox{\columnwidth}{!}{%
    \begin{tabular}{|c|ccc|}
         \hline
         \textbf{\makecell{Preprocessing \\ Approach}}& \textbf{\makecell{Balanced\\Accuracy}} &  \textbf{AUROC}&  \textbf{F1 Score}
\\
         \hline
           Baseline&  52.4\% (49.5-55.1\%)  
& 0.868 (0.847-0.889) & 0.412 (0.380-0.444)
\\
           Reinhard Normalisation& 51.0\% (47.7-54.3\%)   
& 0.870 (0.850-0.888) & 0.392 (0.350-0.437)
\\
           Macenko Normalisation& 45.9\% (41.8-50.0\%)   
& 0.837 (0.814-0.860) & 0.407 (0.360-0.455)
\\
           Otsu Thresholding& \textbf{54.7\%} (51.9-57.6\%)  
& \textbf{0.883} (0.864-0.901) & \textbf{0.440} (0.401-0.482)
\\
           Otsu + Macenko& 44.4\% (40.7-48.3\%)  
& 0.840 (0.816-0.862) & 0.388 (0.347-0.432)
\\
           5x Colour Augmentation& 51.7\% (48.4-54.8\%)  
& 0.867 (0.845-0.887) & 0.401 (0.363-0.441)
\\
           10x Colour Augmentation& 51.1\% (47.8-54.2\%)  
& 0.877 (0.856-0.897) & 0.404 (0.367-0.443)
\\
           20x Colour Augmentation& 51.4\% (48.4-54.4\%)  
& 0.874 (0.853-0.893) & 0.391 (0.352-0.433)
\\
         \hline
    \end{tabular}
    }
    \caption{Results of external validation on the OCEAN Challenge dataset for the ImageNet-pretrained ResNet50 with varied preprocessing approaches, with predictions generated by an ensemble of the five-fold classification models. Results are reported as the mean and 95\% confidence intervals (in brackets) from 10,000 iterations of bootstrapping. The greatest results are shown in \textbf{bold}.}
    \label{table:ocean-results-preprocessing}
\end{table}

\begin{figure}[htbp]
\centering
\includegraphics[width=\textwidth]{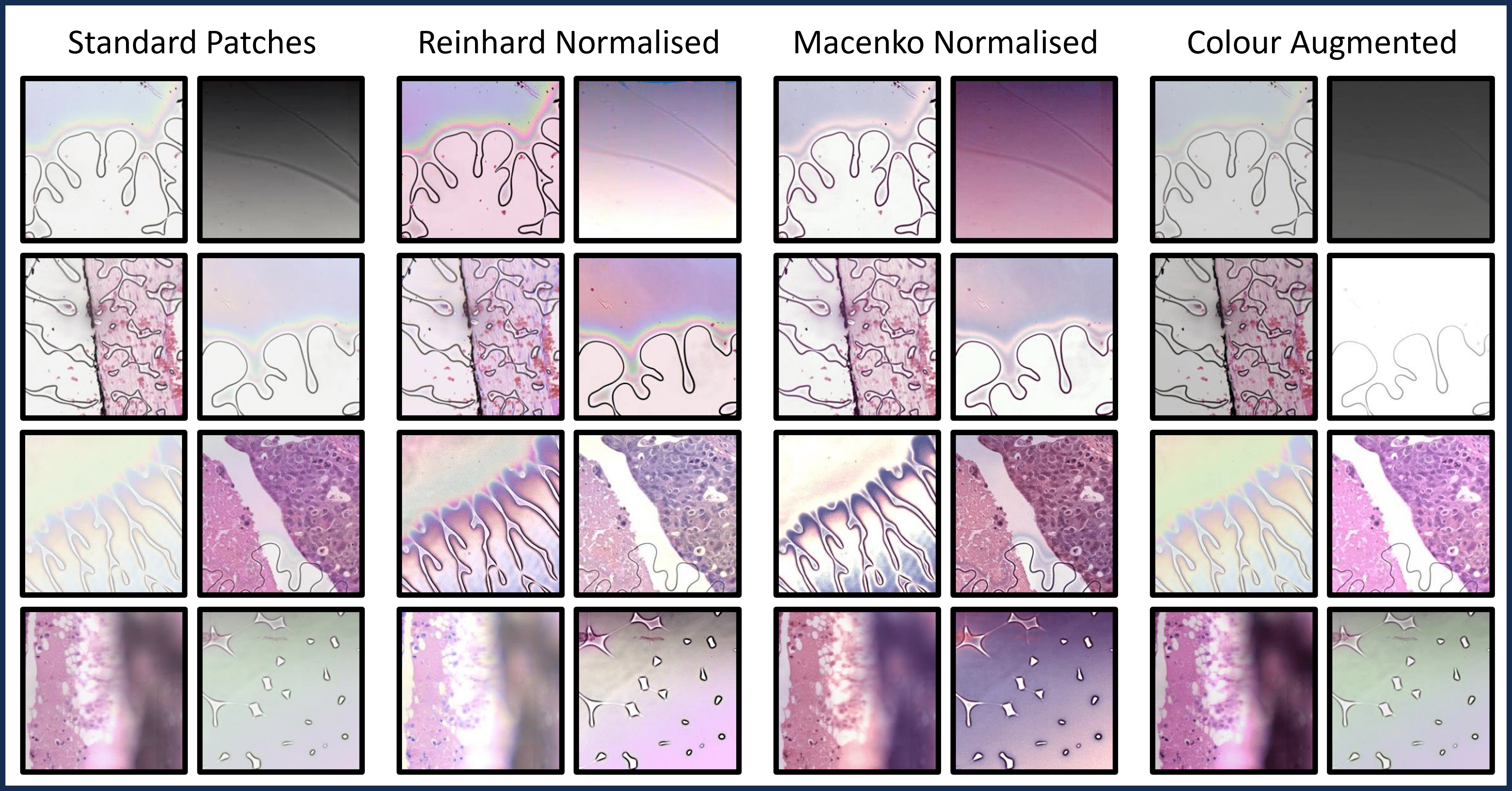}
\caption{Tissue normalisation and augmentation procedures illustrated using 256x256 pixel patches containing artifacts from the same slide at 10x magnification shown in Figure \ref{fig:patchnorm}. The normalisation procedures erroneously apply staining to many non-tissue regions, where the colour augmentations are much less affected. This can be seen most clearly in patches where no tissue is present, such as the bottom-right patch in each group.}
\label{fig:patchnormartif}
\end{figure}

The Reinhard normalisation procedure gave the poorest cross-validation performance. As shown in Figure \ref{fig:patchnormartif}, this method was particularly affected by the presence of artifacts in training slides as the normalisation would make these areas appear similar to tissue, making it harder for the attention mechanism to effectively discard these patches. The impact was smaller on Macenko normalisation, which often avoided applying stain colouring to plain background regions. Colour augmentations were unaffected as they could not introduce staining to non-tissue patches (see Figure \ref{fig:patchnormartif}). 

\begin{figure}[htbp]
\centering
\begin{subfigure}[b]{0.45\textwidth}
\centering
\includegraphics[width=\textwidth]{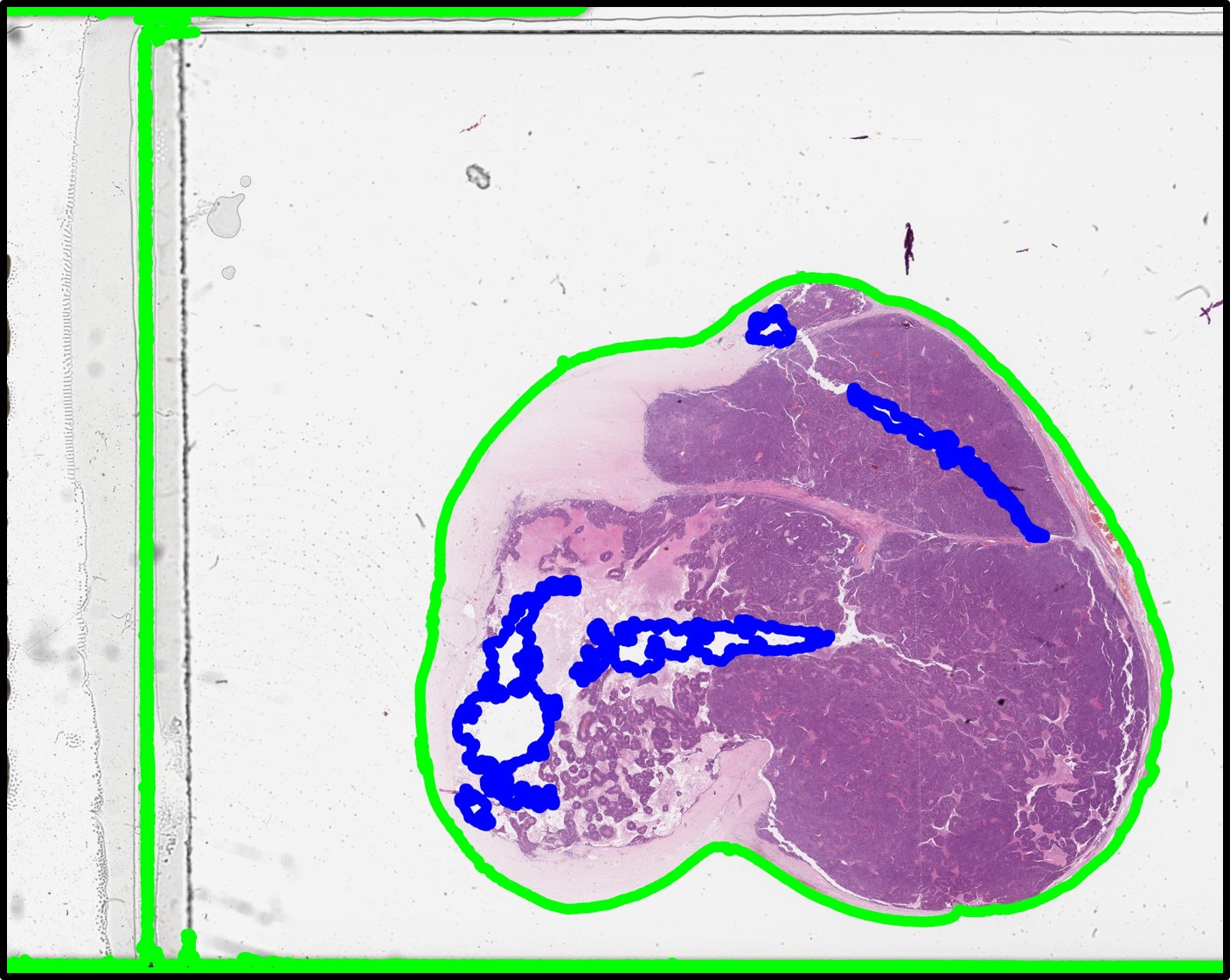}
\caption{Default saturation thresholding \cite{Lu2021}.}
\label{fig:thumbnail}
\end{subfigure}
\hspace{1cm}
\begin{subfigure}[b]{0.45\textwidth}
\centering
\includegraphics[width=\textwidth]{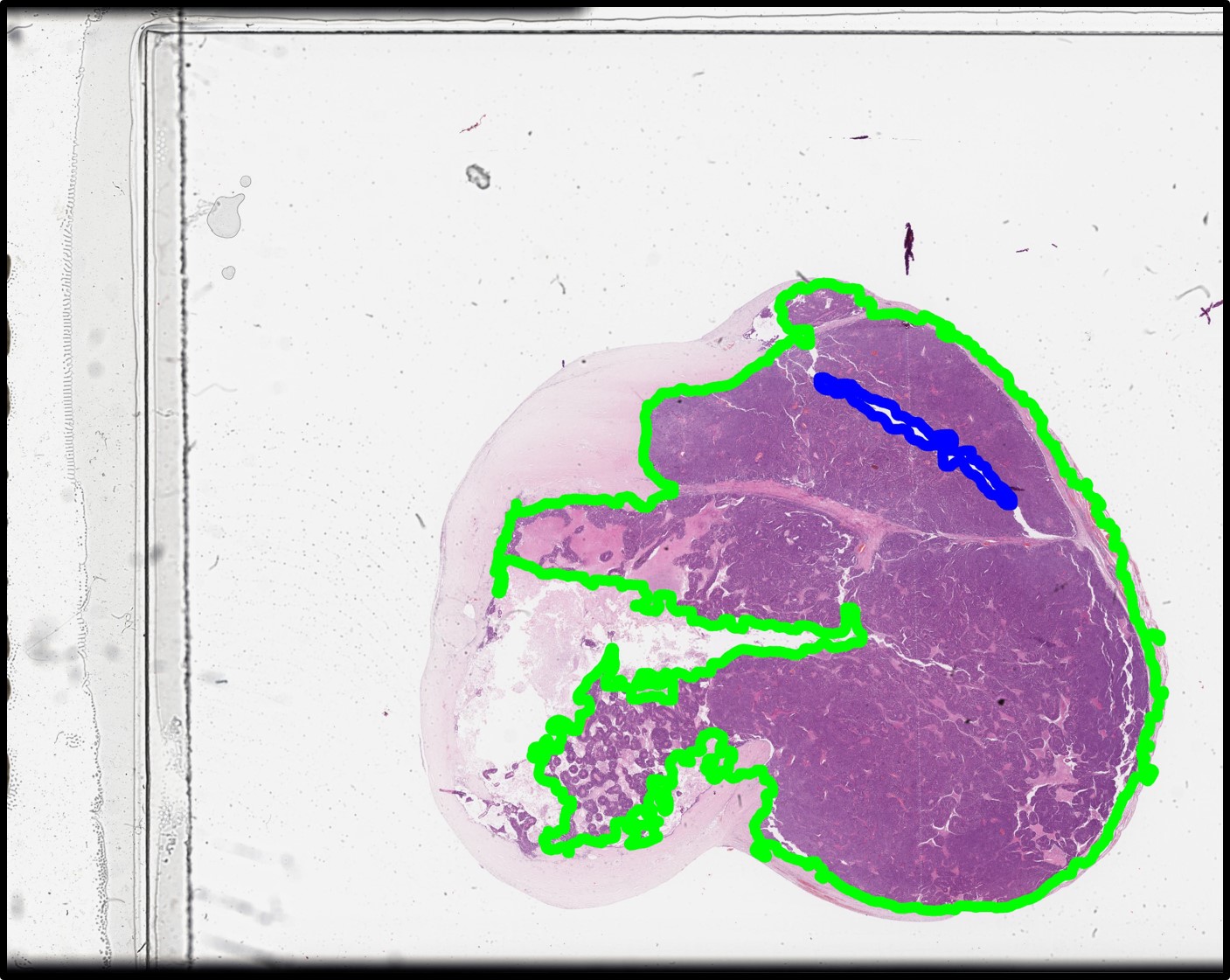}
\caption{Otsu saturation thresholding \cite{Otsu1975}.}
\label{fig:heatmap}
\end{subfigure}
\caption{Examples of two saturation thresholding tissue segmentation approaches, with green outlines indicating tissue and blue outlines indicating holes within the tissue regions. This example contains a coverslip edge which is incorrectly identified as foreground by the default approach, and a small amount of stained tissue which is excluded by Otsu thresholding.}
\label{fig:tissueseg}
\end{figure}

The Otsu procedure was found to remove some unstained tissue and artifacts, while also missing some small areas of stained tissue which may have contained diagnostically relevant information (Figure \ref{fig:tissueseg}). It was unclear from the results whether this was beneficial or harmful overall, with Otsu thresholding giving inconsistent performance when compared to the baseline.

\clearpage
\section{Results of Hypothesis Testing}\label{app:stats}

\begin{table}[h]
    \centering
    \resizebox{0.9\columnwidth}{!}{%
    \begin{tabular}{|c|ccc|ccc|}
        \multicolumn{1}{c}{} & \multicolumn{3}{c}{Cross-Validation p-values} & \multicolumn{3}{c}{Hold-out p-values} \\
       \hline
        Model  & \makecell{Balanced\\Accuracy} &  AUROC& F1 Score & \makecell{Balanced\\Accuracy} &  AUROC& F1 Score \\
    \hline
          RN18 & 0.736 & 0.557 & 0.736 & 0.601 & \textbf{0.028} &  0.317 \\
          ViT-L & \textbf{0.033} & 0.824 & 0.074 & \textbf{0.003} & 0.051 & \textbf{0.002}   \\
          \hline
          RN18-Histo & 0.365 & 0.964 & 0.613 & 0.870 & 0.192 &  0.967 \\
          Lunit & \textbf{0.019} & 0.183 & \textbf{0.009}  & \textbf{0.010} & \textbf{0.035} & \textbf{0.008}  \\
          RN50-Histo & 0.232 & 0.072 & 0.152  & 0.214 & \textbf{0.009} &  0.189 \\
          CTransPath & \textbf{0.009} & 0.082 & \textbf{0.007} & \textbf{0.003} & \textbf{0.012} & \textbf{0.003}   \\
          Hibou-B & \textbf{0.019} & \textbf{0.029} & \textbf{0.012} & \textbf{0.003} & \textbf{0.006} & \textbf{0.003}   \\
          Phikon & \textbf{0.009} & 0.149 & \textbf{0.007} & \textbf{0.003} & \textbf{0.012}  & \textbf{0.003}  \\
          Kaiko-B8 & \textbf{0.013} & 0.063 & \textbf{0.010} & \textbf{0.003} & \textbf{0.011} &  \textbf{0.002} \\
          GPFM & \textbf{0.007} & 0.063 & \textbf{0.007} & \textbf{0.003} & \textbf{0.006}  & \textbf{0.003}  \\
          UNI & \textbf{0.015} & \textbf{0.020} & \textbf{0.009} & \textbf{0.003}  & \textbf{0.006} &  \textbf{0.002} \\
          Hibou-L & \textbf{0.007} & 0.072 & \textbf{0.007}  & \textbf{0.003} & \textbf{0.009} &  \textbf{0.003} \\
          Virchow & \textbf{0.011} & \textbf{0.020} & \textbf{0.006} & \textbf{0.003}  & \textbf{0.006} &  \textbf{0.003} \\
          Virchow2-CLS & \textbf{0.011} & 0.063 & \textbf{0.009} & \textbf{0.002} & \textbf{0.006} & \textbf{0.002}   \\
          H-Optimus-0 & \textbf{0.005} & \textbf{0.020} & \textbf{0.001} & \textbf{0.003}  & \textbf{0.006} &  \textbf{0.002} \\
          Prov-GigaPath & \textbf{0.013} & 0.063 & \textbf{0.007} & \textbf{0.008} & \textbf{0.006} & \textbf{0.008}  \\
    \hline
    
    \multicolumn{7}{c}{} \\
    \multicolumn{1}{c}{} & \multicolumn{3}{c}{Transcanadian p-values} & \multicolumn{3}{c}{UBC-OCEAN p-values} \\
       \hline
        Model  & \makecell{Balanced\\Accuracy} &  AUROC& F1 Score & \makecell{Balanced\\Accuracy} &  AUROC& F1 Score \\
    \hline
          RN18 & 0.446 & 0.773 & 0.399 & 0.237 & \textbf{0.034} & 0.541  \\
          ViT-L & \textbf{0.021} & 0.090 & \textbf{0.022} & 0.170 & 0.265 & \textbf{0.019}  \\
          \hline
          RN18-Histo & 0.490 & 0.211 & 0.403  & 0.235 & 0.987 & \textbf{0.002}  \\
          Lunit & \textbf{0.003} & \textbf{0.011} & \textbf{0.006}  & \textbf{0.002} & \textbf{0.004} & \textbf{0.001}  \\
          RN50-Histo & \textbf{0.015} & \textbf{0.011}  &  \textbf{0.018} & \textbf{0.018} & \textbf{0.004} & \textbf{0.003}  \\
          CTransPath & \textbf{0.007} & \textbf{0.023} & \textbf{0.018} & \textbf{0.002} & \textbf{0.005} & \textbf{0.001}  \\
          Hibou-B & \textbf{0.008} & \textbf{0.024} & \textbf{0.007} & 0.107 & \textbf{0.009} & \textbf{0.019}  \\
          Phikon & \textbf{0.002} & \textbf{0.011} & \textbf{$<$0.001} & \textbf{0.001} & \textbf{0.013} &  \textbf{0.001} \\
          Kaiko-B8 & \textbf{0.007} & \textbf{0.011} & \textbf{0.022}  & \textbf{0.004}  & \textbf{0.006} & \textbf{0.003}  \\
          GPFM & \textbf{0.003} & \textbf{0.011}  & \textbf{0.006} & \textbf{0.001} & \textbf{0.004} &  \textbf{0.001} \\
          UNI & \textbf{0.006} & \textbf{0.011} & \textbf{0.015} & \textbf{0.001} & \textbf{0.004} & \textbf{0.001}  \\
          Hibou-L & \textbf{0.003} & \textbf{0.013} & \textbf{0.003}  & \textbf{0.002} & \textbf{0.005} & \textbf{0.002}  \\
          Virchow & \textbf{0.006} & \textbf{0.018}  & \textbf{0.015} & \textbf{0.002} & \textbf{0.004} & \textbf{0.001}  \\
          Virchow2-CLS & \textbf{0.007} & \textbf{0.011}  & \textbf{0.015} & \textbf{0.001} & \textbf{0.004} & \textbf{0.001}   \\
          H-Optimus-0 & \textbf{0.003} & \textbf{0.011} & \textbf{0.004} & \textbf{0.002} & \textbf{0.004} & \textbf{0.001}  \\
          Prov-GigaPath & \textbf{0.019} & \textbf{0.017} & \textbf{0.035} & \textbf{0.005} & \textbf{0.004} &  \textbf{0.002} \\
    \hline
 \end{tabular}
 }
    \caption{Resulting p-values from paired t-tests comparing the subtype classification results with each feature extractor to the ImageNet-pretrained ResNet50 baseline. False discovery rate p-value adjustments were applied to account for multiple testing \cite{Benjamini1995}. Values below 0.05 are indicated in \textbf{bold}.}
    \label{tab:pval_baseline}
\end{table}

\begin{table}[h]
    \centering
    \resizebox{0.9\columnwidth}{!}{%
    \begin{tabular}{|c|ccc|ccc|}
        \multicolumn{1}{c}{} & \multicolumn{3}{c}{Cross-Validation p-values} & \multicolumn{3}{c}{Hold-out p-values} \\
       \hline
        Model  & \makecell{Balanced\\Accuracy} &  AUROC& F1 Score & \makecell{Balanced\\Accuracy} &  AUROC& F1 Score \\
    \hline
          RN50 & 0.617 & 0.264 & 0.133 & 0.171 & 0.252 & 0.133    \\  
          RN18 & 0.967 & 0.259 & 0.170 & 0.326  & 0.252 & 0.170     \\
          ViT-L & \textbf{0.012} & \textbf{0.005} & \textbf{0.010} & \textbf{0.006} & 0.095 & \textbf{0.010}    \\
          \hline
          RN18-Histo & \textbf{0.002} & 0.095 &  0.145  & 0.086 & 0.671 & 0.145   \\
          Lunit & 0.555 & \textbf{0.011} & 0.168   & 0.054 & 0.124  &  0.074 \\
          RN50-Histo & 0.864 & 0.630 &  0.902 & 0.912 & 0.100 & 0.895  \\
          CTransPath & 0.144 & 0.987 & \textbf{0.042}   & \textbf{0.030}  & 0.099 & \textbf{0.042}  \\
          Hibou-B & 0.159 & 0.069 & \textbf{0.009} & \textbf{0.008} & 0.207   & \textbf{0.009}   \\
          Phikon & 0.709 & 0.280 & 0.741 & 0.619 & 0.114    & 0.741  \\
          Kaiko-B8 & \textbf{0.039} & 0.089 & 0.124   & 0.099 & 0.063 & 0.124  \\
          GPFM & 0.500 & \textbf{0.029} & 0.236 & 0.262 & 0.055  & 0.236    \\
          UNI & \textbf{0.003} & \textbf{0.015} & \textbf{0.021}  & \textbf{0.033} & 0.614 & \textbf{0.021}    \\
          Hibou-L & 0.104 & 0.050 &  0.070 & \textbf{0.019} & 0.193 & \textbf{0.016}  \\
          Virchow & \textbf{0.039} & 0.059 & 0.104 & 0.069  & 0.076   & 0.104  \\
          Virchow2-CLS & 0.194 & \textbf{0.035} & 0.095 & 0.083  & 0.108  & 0.095    \\
          H-Optimus-0 & 0.111 & 0.069 & 0.133  & 0.119 & 0.089 & 0.133     \\
          Prov-GigaPath & 0.412 & 0.297  & 0.215 & 0.194 & \textbf{0.035} & 0.215    \\
    \hline
        \multicolumn{7}{c}{} \\
        \multicolumn{1}{c}{} & \multicolumn{3}{c}{Transcanadian p-values} & \multicolumn{3}{c}{UBC-OCEAN p-values} \\
       \hline
        Model  & \makecell{Balanced\\Accuracy} &  AUROC& F1 Score & \makecell{Balanced\\Accuracy} &  AUROC& F1 Score \\
    \hline
          RN50 & 0.190 & 0.178 & 0.219 & 0.303 & 0.098 & 0.716   \\    
          RN18 & 0.240 & 0.217 & 0.106 & 0.339 & 0.056 & 0.279  \\
          ViT-L & \textbf{0.014} & 0.109 & \textbf{0.014} & \textbf{0.001} & \textbf{0.006} & \textbf{0.021}  \\
          \hline
          RN18-Histo & 0.578 & 0.774 & 0.973 & 0.212  & 0.182 & 0.620  \\
          Lunit & 0.099 & 0.774 & 0.099  & 0.104 & \textbf{0.049} & 0.056  \\
          RN50-Histo & 0.601 & 0.135 & 0.479  & 0.818 & \textbf{0.023} & 0.605  \\
          CTransPath & 0.853 & 0.341 & 0.998 & 0.790 & 0.630  &  0.562 \\
          Hibou-B & 0.300 & 0.076 & 0.286 & 0.700 & 0.172 &  0.590 \\
          Phikon & 0.740  & 0.085 & 0.306 & 0.119 & 0.467 &  0.189 \\
          Kaiko-B8 & 0.213 & 0.125  & 0.342 & 0.102 & \textbf{0.014} &  \textbf{0.028} \\
          GPFM & 0.386 & 0.120 & 0.405  & 0.861 & 0.080 & 0.176  \\
          UNI & 0.370 & 0.085 & 0.959 & \textbf{0.002} & \textbf{0.008} & \textbf{0.014}  \\
          Hibou-L & 0.087 & \textbf{0.040} & \textbf{0.003}  & 0.142 & \textbf{0.004} &  0.196 \\
          Virchow & 0.478 & 0.379 & 0.460 & \textbf{0.049} & \textbf{0.012} & \textbf{0.057}  \\
          Virchow2-CLS & 0.057 & 0.871 & 0.057 & \textbf{0.017} & 0.066 &  \textbf{0.035} \\
          H-Optimus-0 & 0.167 & 0.751 & 0.192  & 0.866 & 0.054 &  0.168 \\
          Prov-GigaPath & 0.274 & 0.060  & 0.209  & 0.258 & 0.124 & 0.416  \\
    \hline
 \end{tabular}
 }
    \caption{Resulting p-values from paired t-tests comparing the subtype classification results for each feature extractor with and without hyperparameter tuning applied to the ABMIL classifier. Values below 0.05 are indicated in \textbf{bold}.}
    \label{tab:pval_hyperparams}
\end{table}

\clearpage

\section{Supplementary Attention Heatmap Analysis}\label{app:heatmaps}
\begin{figure}[htbp]
\centering
\includegraphics[width=\textwidth]{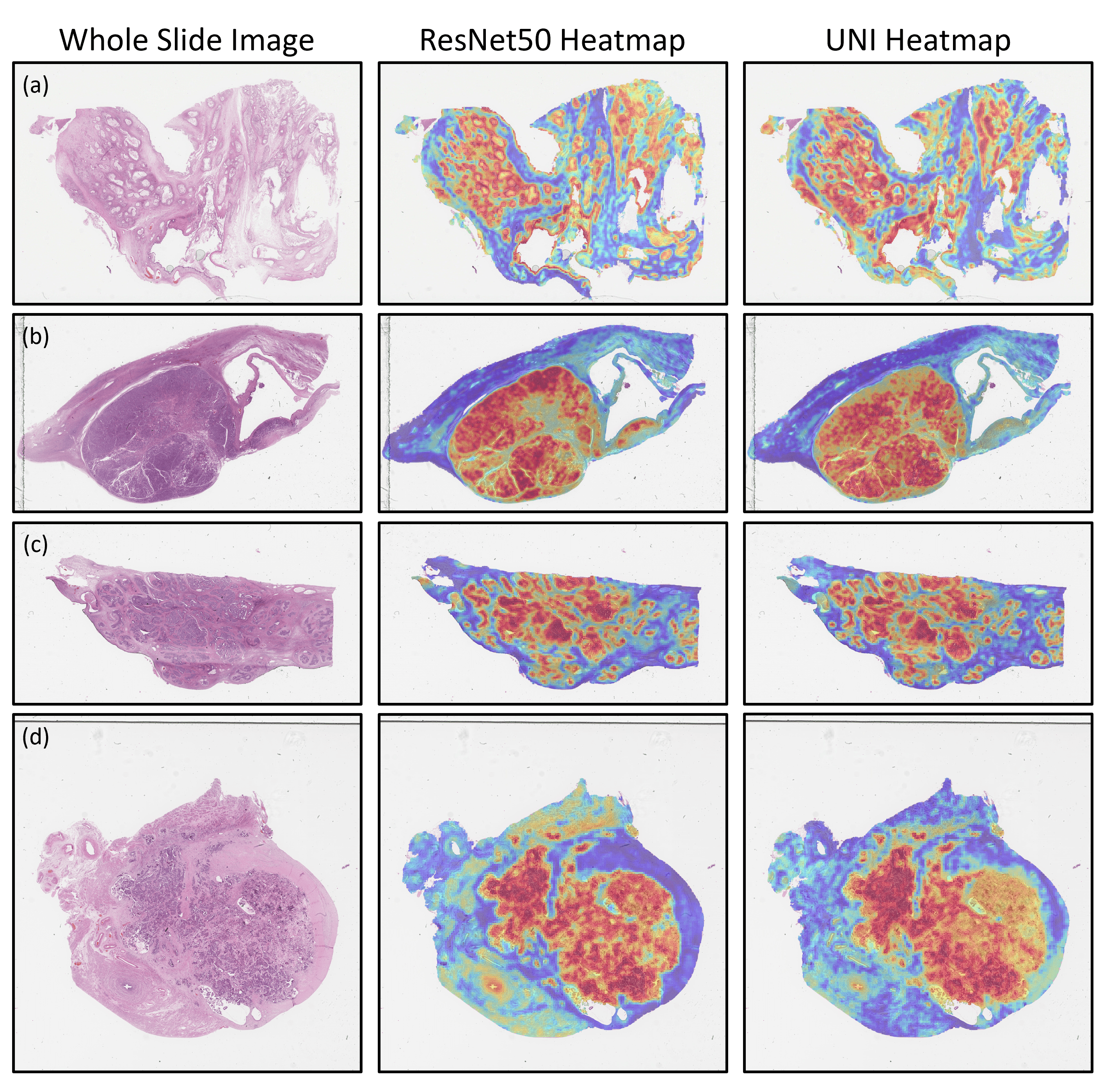}
\caption{Attention heatmaps from the ABMIL classifier using ImageNet-pretrained ResNet50 and UNI foundation model features, where the classification differed between the two models. (a) Ground truth: MC, ResNet50: CCC, UNI: MC. (b) Ground truth: CCC, ResNet50: HGSC, UNI: CCC. (c) Ground truth: EC, ResNet50: HGSC, UNI: EC. (d) Ground truth: LGSC, ResNet50: HGSC, UNI: LGSC.}
\label{fig:appheatmaps1}
\end{figure}

\begin{figure}[htbp]
\centering
\includegraphics[width=\textwidth]{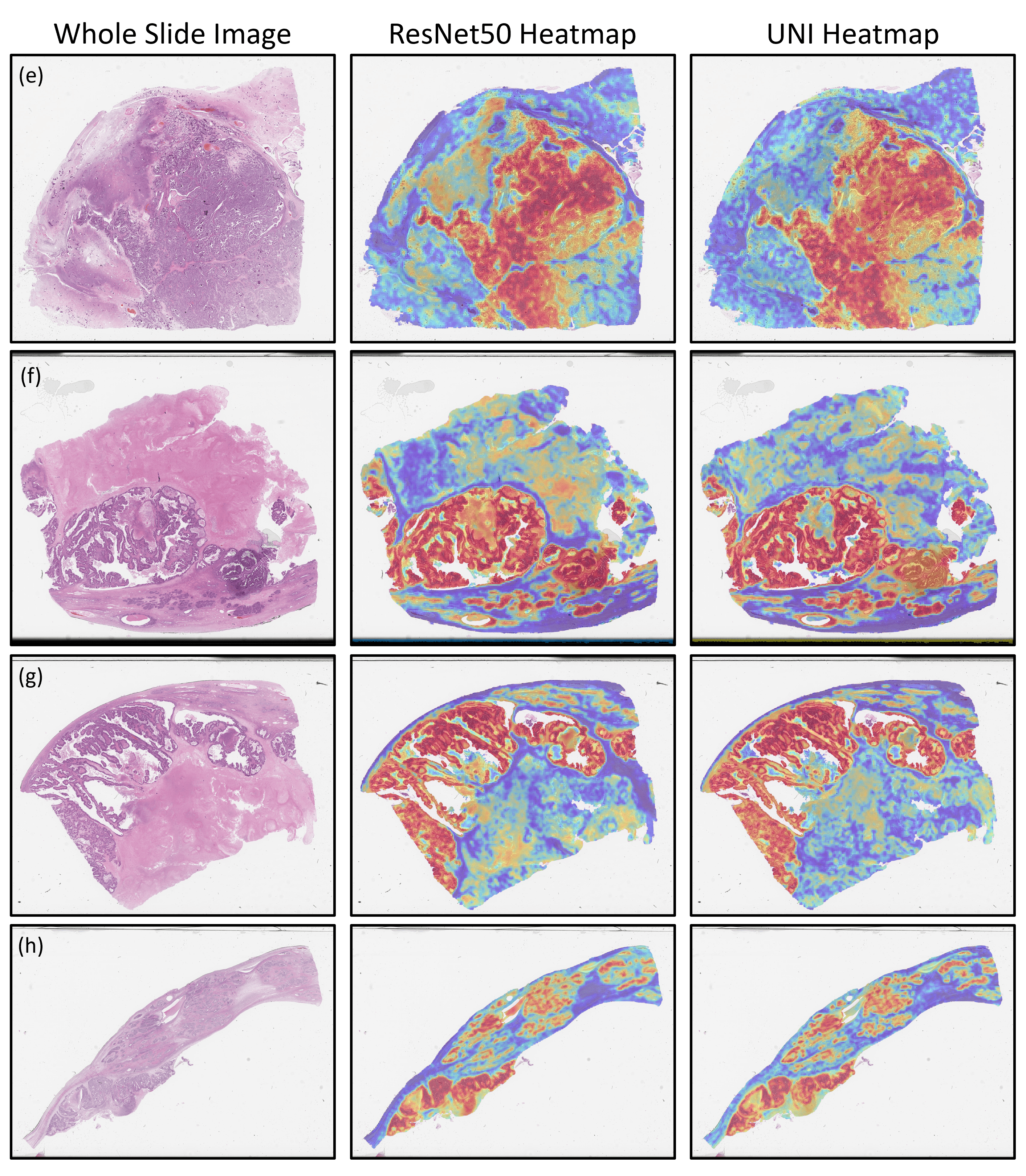}
\caption{Attention heatmaps from the ABMIL classifier using ImageNet-pretrained ResNet50 and UNI foundation model features, where the classification differed between the two models. (e) Ground truth: LGSC, ResNet50: HGSC, UNI: LGSC. (f) Ground truth: HGSC, ResNet50: HGSC, UNI: EC. (g) Ground truth: HGSC, ResNet50: HGSC, UNI: EC. (h) Ground truth: EC, ResNet50: HGSC, UNI: EC.}
\label{fig:appheatmaps2}
\end{figure}

Two pathologists (KA and NMO) qualitatively compared the UNI and ImageNet-pretrained attention heatmaps for ten class-balanced example WSIs from the internal hold-out test set. These WSIs (shown in Figures \ref{fig:heatmaps}, \ref{fig:appheatmaps1}, and \ref{fig:appheatmaps2}) were selected from those in which a different classification had been determined by each model (specifically using the first-fold model of the five-model ensemble).  Out of 39 total disagreements, the UNI-based model gave the correct classification in 26 cases, the ResNet50-based model in 3 cases, and neither was correct in 10 cases. The pathologists were only provided the heatmaps, and were blinded to the models used and the predictions made.    

The heatmaps were determined to be similar between models, with both giving high attention to tumour regions and low attention to most stroma regions. Where differences occurred, the ResNet50-based model would typically give high attention to a larger tissue area, often including relevant stromal features (e.g. necrosis and psammoma bodies), but sometimes also including irrelevant stroma. When considering whether heatmaps had focused on diagnostically relevant regions, the pathologists expressed a preference for the UNI-based heatmap in four cases and the ResNet50-based heatmap in three cases, with no preference expressed for the remaining three cases due to their overwhelming similarity. In eight of the selected cases, the UNI model had correctly determined the classification, including all three cases in which the pathologists had preferred the ResNet50-based heatmap. In these cases, the UNI model did not appear to give sufficient attention to all relevant tissue, though it still determined the correct classification. Thus, there was some level of divergence between the pathologists' interpretations and the model heatmaps.        

\clearpage
\section{TRIPOD+AI Reporting Checklist}\label{app:tripod}

\begin{figure}[h] \centering{
\fbox{\includegraphics[width=0.75\textwidth,page=1]{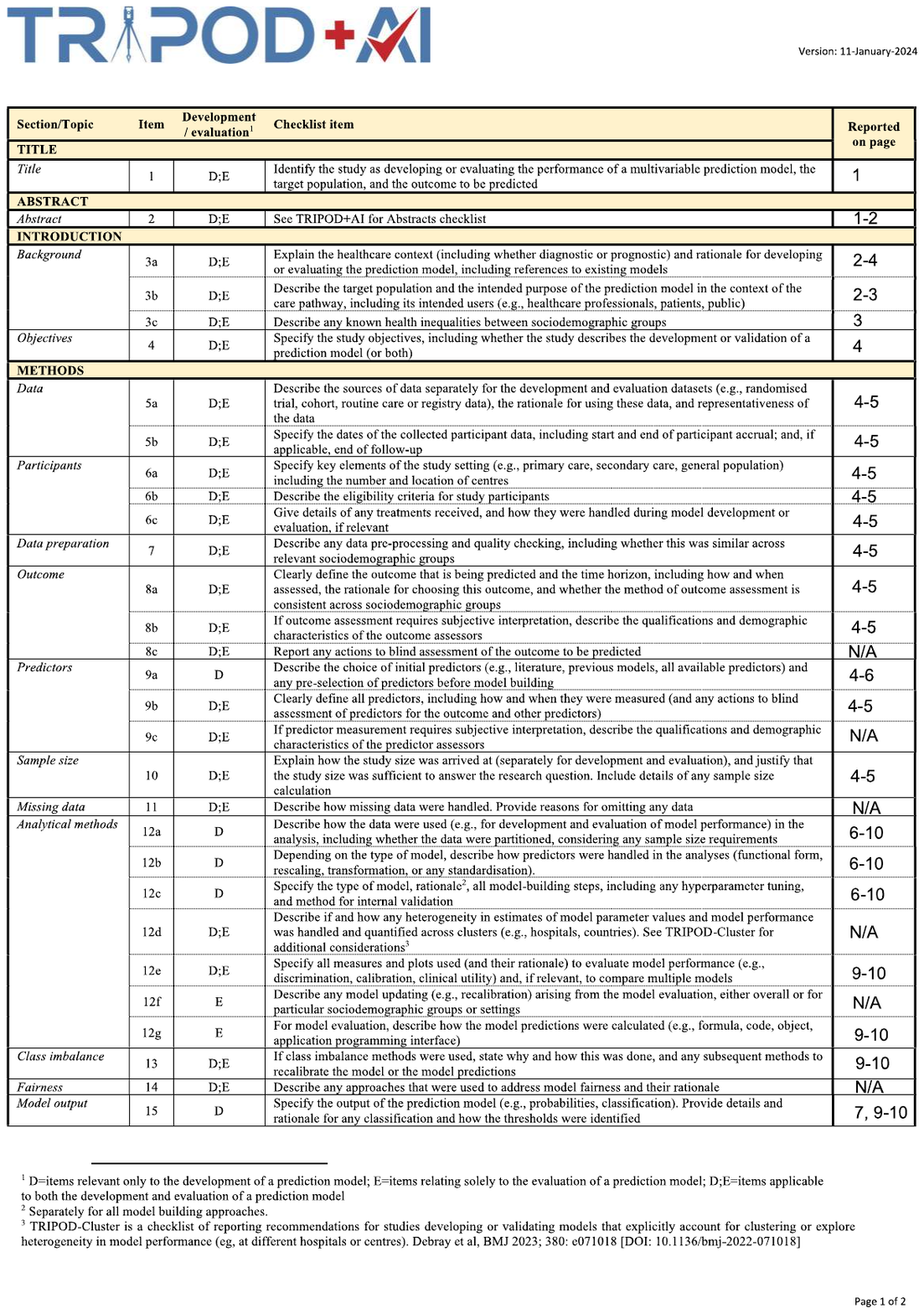}}}
\end{figure} 
\newpage
\begin{figure}[h] \centering{
\fbox{\includegraphics[width=0.75\textwidth,page=2]{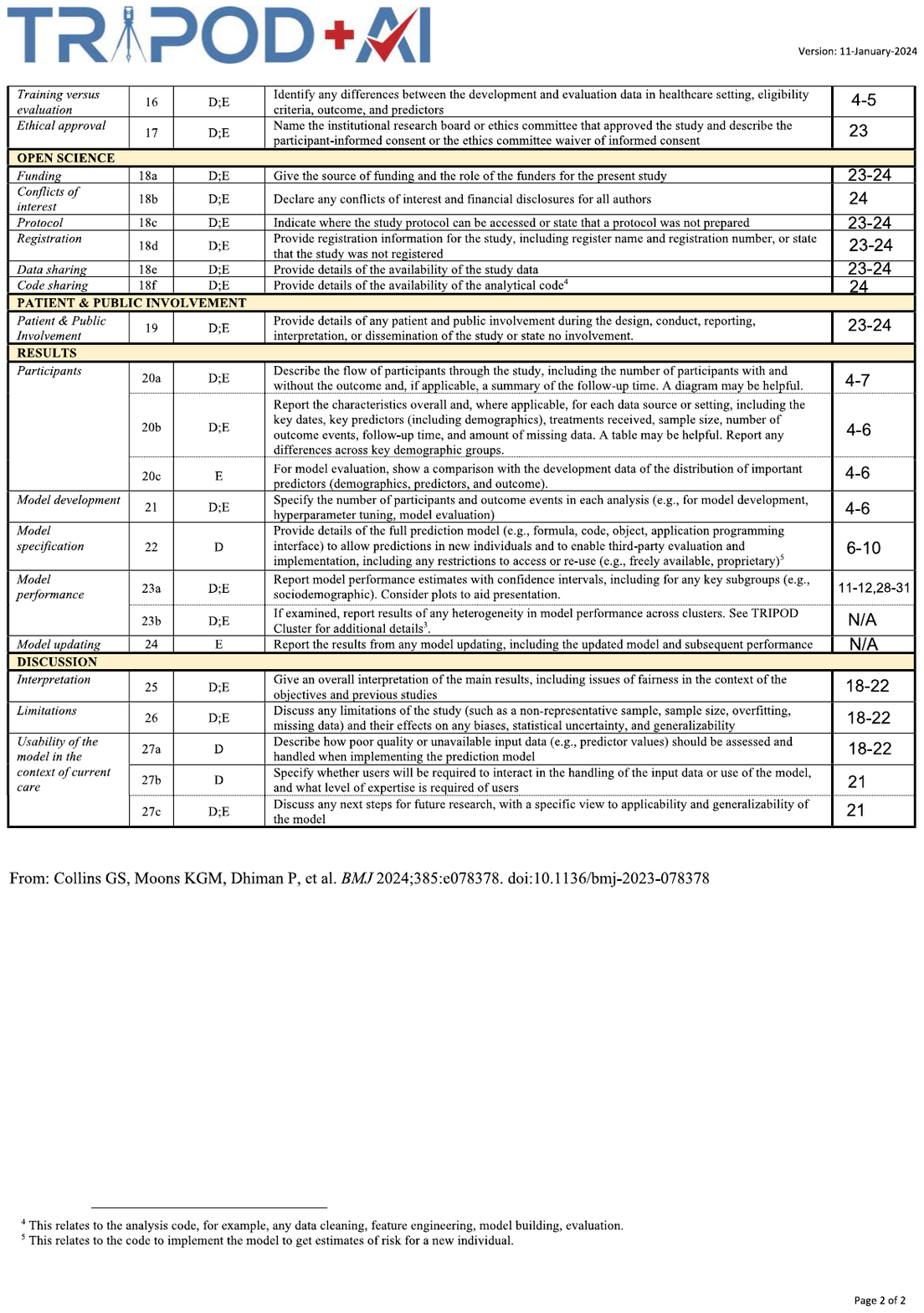}}}
\end{figure} 

\clearpage





\bibliographystyle{elsarticle-num}
\bibliography{main}

\begin{thebibliography}{10}
\expandafter\ifx\csname url\endcsname\relax
  \def\url#1{\texttt{#1}}\fi
\expandafter\ifx\csname urlprefix\endcsname\relax\def\urlprefix{URL }\fi
\expandafter\ifx\csname href\endcsname\relax
  \def\href#1#2{#2} \def\path#1{#1}\fi

\bibitem{Bray2024}
F.~Bray, M.~Laversanne, H.~Sung, J.~Ferlay, R.~L. Siegel, I.~Soerjomataram, A.~Jemal, Global cancer statistics 2022: Globocan estimates of incidence and mortality worldwide for 36 cancers in 185 countries, CA: A Cancer Journal for Clinicians (2024) 1--35\href {https://doi.org/https://doi.org/10.3322/caac.21834} {\path{doi:https://doi.org/10.3322/caac.21834}}.

\bibitem{Kobel2008}
M.~K{\"o}bel, S.~E. Kalloger, N.~Boyd, S.~McKinney, E.~Mehl, C.~Palmer, S.~Leung, N.~J. Bowen, D.~N. Ionescu, A.~Rajput, L.~M. Prentice, D.~Miller, J.~Santos, K.~Swenerton, C.~B. Gilks, D.~Huntsman, Ovarian carcinoma subtypes are different diseases: implications for biomarker studies, PLoS medicine 5~(12) (2008) e232.

\bibitem{Peres2019}
L.~C. Peres, K.~L. Cushing-Haugen, M.~K{\"o}bel, H.~R. Harris, A.~Berchuck, M.~A. Rossing, J.~M. Schildkraut, J.~A. Doherty, Invasive epithelial ovarian cancer survival by histotype and disease stage, JNCI: Journal of the National Cancer Institute 111~(1) (2019) 60--68.

\bibitem{Moch2020}
H.~Moch, Female genital tumours: {WHO} classification of tumours, volume 4, WHO Classification of Tumours 4 (2020).

\bibitem{Vroobel2024}
K.~Vroobel, Overview of ovarian tumours: Pathogenesis and general considerations, in: Pathology of the Ovary, Fallopian Tube and Peritoneum, Springer, 2024, pp. 95--113.

\bibitem{Kobel2014}
M.~Köbel, J.~Bak, B.~I. Bertelsen, O.~Carpen, A.~Grove, E.~S. Hansen, A.-M. Levin~Jakobsen, M.~Lidang, A.~Måsbäck, A.~Tolf, C.~B. Gilks, J.~W. Carlson, Ovarian carcinoma histotype determination is highly reproducible, and is improved through the use of immunohistochemistry, Histopathology 64~(7) (2014) 1004--1013.

\bibitem{RCPath2018}
{Royal College of Pathologists}, \href{https://www.rcpath.org/static/952a934d-2ec3-48c9-a8e6e00fcdca700f/Meeting-Pathology-Demand-Histopathology-Workforce-Census-2018.pdf}{Meeting pathology demand: Histopathology workforce census} (2018).
\newline\urlprefix\url{https://www.rcpath.org/static/952a934d-2ec3-48c9-a8e6e00fcdca700f/Meeting-Pathology-Demand-Histopathology-Workforce-Census-2018.pdf}

\bibitem{Wilson2018}
M.~L. Wilson, K.~A. Fleming, M.~A. Kuti, L.~M. Looi, N.~Lago, K.~Ru, Access to pathology and laboratory medicine services: a crucial gap, The Lancet 391~(10133) (2018) 1927--1938.

\bibitem{Hanna2020}
T.~P. Hanna, W.~D. King, S.~Thibodeau, M.~Jalink, G.~A. Paulin, E.~Harvey-Jones, D.~E. O’Sullivan, C.~M. Booth, R.~Sullivan, A.~Aggarwal, Mortality due to cancer treatment delay: systematic review and meta-analysis, bmj 371 (2020).

\bibitem{Allen2024}
K.~E. Allen, P.~Adusumilli, J.~Breen, G.~Hall, N.~M. Orsi, Artificial intelligence in ovarian digital pathology, in: Pathology of the Ovary, Fallopian Tube and Peritoneum, Springer, 2024, pp. 731--749.

\bibitem{Breen2023review}
J.~Breen, K.~Allen, K.~Zucker, P.~Adusumilli, A.~Scarsbrook, G.~Hall, N.~M. Orsi, N.~Ravikumar, Artificial intelligence in ovarian cancer histopathology: a systematic review, NPJ Precision Oncology 7~(1) (2023) 83.

\bibitem{Matthews2024}
G.~A. Matthews, C.~McGenity, D.~Bansal, D.~Treanor, Public evidence on ai products for digital pathology, medRxiv (2024) 2024--02.

\bibitem{Breen2023}
J.~Breen, K.~Allen, K.~Zucker, G.~Hall, N.~M. Orsi, N.~Ravikumar, {Efficient subtyping of ovarian cancer histopathology whole slide images using active sampling in multiple instance learning}, in: J.~E. Tomaszewski, A.~D. Ward (Eds.), Medical Imaging 2023: Digital and Computational Pathology, Vol. 12471, International Society for Optics and Photonics, SPIE, 2023, p. 1247110.
\newblock \href {https://doi.org/10.1117/12.2653869} {\path{doi:10.1117/12.2653869}}.

\bibitem{Breen2024ISBI}
J.~Breen, K.~Allen, K.~Zucker, N.~M. Orsi, N.~Ravikumar, Reducing histopathology slide magnification improves the accuracy and speed of ovarian cancer subtyping, arXiv preprint arXiv:2311.13956 (2023).

\bibitem{Breen2024graph}
J.~Breen, K.~Allen, K.~Zucker, N.~M. Orsi, N.~Ravikumar, Multi-resolution histopathology patch graphs for ovarian cancer subtyping, arXiv preprint arXiv:2407.18105 (2024).

\bibitem{BenTaieb2015}
A.~BenTaieb, H.~Li-Chang, D.~Huntsman, G.~Hamarneh, Automatic diagnosis of ovarian carcinomas via sparse multiresolution tissue representation, in: Medical Image Computing and Computer-Assisted Intervention--MICCAI 2015: 18th International Conference, Munich, Germany, October 5-9, 2015, Proceedings, Part I 18, Springer, 2015, pp. 629--636.

\bibitem{BenTaieb2016}
A.~BenTaieb, M.~S. Nosrati, H.~Li-Chang, D.~Huntsman, G.~Hamarneh, Clinically-inspired automatic classification of ovarian carcinoma subtypes, Journal of pathology informatics 7~(1) (2016) 28.

\bibitem{BenTaieb2017}
A.~BenTaieb, H.~Li-Chang, D.~Huntsman, G.~Hamarneh, A structured latent model for ovarian carcinoma subtyping from histopathology slides, Medical image analysis 39 (2017) 194--205.

\bibitem{Levine2020}
A.~B. Levine, J.~Peng, D.~Farnell, M.~Nursey, Y.~Wang, J.~R. Naso, H.~Ren, H.~Farahani, C.~Chen, D.~Chiu, A.~Talhouk, B.~Sheffield, M.~Riazy, P.~P. Ip, C.~Parra-Herran, A.~Mills, N.~Singh, B.~Tessier-Cloutier, T.~Salisbury, J.~Lee, T.~Salcudean, S.~J. Jones, D.~G. Huntsman, C.~B. Gilks, S.~Yip, A.~Bashashati, Synthesis of diagnostic quality cancer pathology images by generative adversarial networks, The Journal of pathology 252~(2) (2020) 178--188.

\bibitem{Boschman2022}
J.~Boschman, H.~Farahani, A.~Darbandsari, P.~Ahmadvand, A.~Van~Spankeren, D.~Farnell, A.~B. Levine, J.~R. Naso, A.~Churg, S.~J. Jones, S.~Yip, M.~Köbel, D.~G. Huntsman, C.~B. Gilks, A.~Bashashati, The utility of color normalization for ai-based diagnosis of hematoxylin and eosin-stained pathology images, The Journal of Pathology 256~(1) (2022) 15--24.

\bibitem{Farahani2022}
H.~Farahani, J.~Boschman, D.~Farnell, A.~Darbandsari, A.~Zhang, P.~Ahmadvand, S.~J. Jones, D.~Huntsman, M.~Köbel, C.~B. Gilks, N.~Singh, A.~Bashashati, Deep learning-based histotype diagnosis of ovarian carcinoma whole-slide pathology images, Modern Pathology 35~(12) (2022) 1983--1990.

\bibitem{Mirabadi2024}
A.~K. Mirabadi, G.~Archibald, A.~Darbandsari, A.~Contreras-Sanz, R.~E. Nakhli, M.~Asadi, A.~Zhang, C.~B. Gilks, P.~Black, G.~Wang, H.~Farahani, A.~Bashashati, {GRASP}: Graph-structured pyramidal whole slide image representation, arXiv preprint arXiv:2402.03592 (2024).

\bibitem{Asadi2024}
M.~Asadi-Aghbolaghi, A.~Darbandsari, A.~Zhang, A.~Contreras-Sanz, J.~Boschman, P.~Ahmadvand, M.~Köbel, D.~Farnell, D.~G. Huntsman, A.~Churg, P.~C. Black, G.~Wang, C.~B. Gilks, H.~Farahani, A.~Bashashati, Learning generalizable ai models for multi-center histopathology image classification, npj Precision Oncology 8~(1) (2024) 151.

\bibitem{Ma2024}
J.~Ma, Z.~Guo, F.~Zhou, Y.~Wang, Y.~Xu, Y.~Cai, Z.~Zhu, C.~Jin, Y.~Lin, X.~Jiang, A.~Han, L.~Liang, R.~C.~K. Chan, J.~Wang, K.-T. Cheng, H.~Chen, Towards a generalizable pathology foundation model via unified knowledge distillation, arXiv preprint arXiv:2407.18449 (2024).

\bibitem{Xu2024}
H.~Xu, N.~Usuyama, J.~Bagga, S.~Zhang, R.~Rao, T.~Naumann, C.~Wong, Z.~Gero, J.~González, Y.~Gu, Y.~Xu, M.~Wei, W.~Wang, S.~Ma, F.~Wei, J.~Yang, C.~Li, J.~Gao, J.~Rosemon, T.~Bower, S.~Lee, R.~Weerasinghe, B.~J. Wright, A.~Robicsek, B.~Piening, C.~Bifulco, S.~Wang, H.~Poon, A whole-slide foundation model for digital pathology from real-world data, Nature (2024) 1--8.

\bibitem{Gadermayr2024}
M.~Gadermayr, M.~Tschuchnig, Multiple instance learning for digital pathology: A review of the state-of-the-art, limitations \& future potential, Computerized Medical Imaging and Graphics (2024) 102337.

\bibitem{He2016}
K.~He, X.~Zhang, S.~Ren, J.~Sun, Deep residual learning for image recognition, in: Proceedings of the IEEE conference on computer vision and pattern recognition, 2016, pp. 770--778.

\bibitem{Lu2021}
M.~Y. Lu, D.~F. Williamson, T.~Y. Chen, R.~J. Chen, M.~Barbieri, F.~Mahmood, Data-efficient and weakly supervised computational pathology on whole-slide images, Nature biomedical engineering 5~(6) (2021) 555--570.

\bibitem{Shao2021}
Z.~Shao, H.~Bian, Y.~Chen, Y.~Wang, J.~Zhang, X.~Ji, Transmil: Transformer based correlated multiple instance learning for whole slide image classification, Advances in neural information processing systems 34 (2021) 2136--2147.

\bibitem{Zaffar2023}
I.~Zaffar, G.~Jaume, N.~Rajpoot, F.~Mahmood, Embedding space augmentation for weakly supervised learning in whole-slide images, in: 2023 IEEE 20th International Symposium on Biomedical Imaging (ISBI), IEEE, 2023, pp. 1--4.

\bibitem{Godson2024}
L.~Godson, N.~Alemi, J.~Nsengimana, G.~P. Cook, E.~L. Clarke, D.~Treanor, D.~T. Bishop, J.~Newton-Bishop, A.~Gooya, D.~Magee, Immune subtyping of melanoma whole slide images using multiple instance learning, Medical Image Analysis 93 (2024) 103097.

\bibitem{Russakovsky2015}
O.~Russakovsky, J.~Deng, H.~Su, J.~Krause, S.~Satheesh, S.~Ma, Z.~Huang, A.~Karpathy, A.~Khosla, M.~Bernstein, A.~C. Berg, L.~Fei-Fei, Imagenet large scale visual recognition challenge, International journal of computer vision 115 (2015) 211--252.

\bibitem{Kobel2010}
M.~Köbel, S.~E. Kalloger, P.~M. Baker, C.~A. Ewanowich, J.~Arseneau, V.~Zherebitskiy, S.~Abdulkarim, S.~Leung, M.~A. Duggan, D.~Fontaine, R.~Parker, D.~G. Huntsman, C.~B. Gilks, Diagnosis of ovarian carcinoma cell type is highly reproducible: a transcanadian study, The American journal of surgical pathology 34~(7) (2010) 984--993.

\bibitem{Ciga2022}
O.~Ciga, T.~Xu, A.~L. Martel, Self supervised contrastive learning for digital histopathology, Machine Learning with Applications 7 (2022) 100198.

\bibitem{Chen2022}
R.~J. Chen, C.~Chen, Y.~Li, T.~Y. Chen, A.~D. Trister, R.~G. Krishnan, F.~Mahmood, Scaling vision transformers to gigapixel images via hierarchical self-supervised learning, in: Proceedings of the IEEE/CVF Conference on Computer Vision and Pattern Recognition, 2022, pp. 16144--16155.

\bibitem{Wang2022}
X.~Wang, S.~Yang, J.~Zhang, M.~Wang, J.~Zhang, W.~Yang, J.~Huang, X.~Han, Transformer-based unsupervised contrastive learning for histopathological image classification, Medical image analysis 81 (2022) 102559.

\bibitem{Kang2023}
M.~Kang, H.~Song, S.~Park, D.~Yoo, S.~Pereira, Benchmarking self-supervised learning on diverse pathology datasets, in: Proceedings of the IEEE/CVF Conference on Computer Vision and Pattern Recognition, 2023, pp. 3344--3354.

\bibitem{Filiot2023}
A.~Filiot, R.~Ghermi, A.~Olivier, P.~Jacob, L.~Fidon, A.~Mac~Kain, C.~Saillard, J.-B. Schiratti, Scaling self-supervised learning for histopathology with masked image modeling, medRxiv (2023) 2023--07.

\bibitem{Wang2023}
W.~Wang, S.~Ma, H.~Xu, N.~Usuyama, J.~Ding, H.~Poon, F.~Wei, When an image is worth 1,024 x 1,024 words: A case study in computational pathology, arXiv preprint arXiv:2312.03558 (2023).

\bibitem{Nechaev2024}
D.~Nechaev, A.~Pchelnikov, E.~Ivanova, Hibou: A family of foundational vision transformers for pathology, arXiv preprint arXiv:2406.05074 (2024).

\bibitem{Vorontsov2024}
E.~Vorontsov, A.~Bozkurt, A.~Casson, G.~Shaikovski, M.~Zelechowski, K.~Severson, E.~Zimmermann, J.~Hall, N.~Tenenholtz, N.~Fusi, E.~Yang, P.~Mathieu, A.~van Eck, D.~Lee, J.~Viret, E.~Robert, Y.~K. Wang, J.~D. Kunz, M.~C.~H. Lee, J.~H. Bernhard, R.~A. Godrich, G.~Oakley, E.~Millar, M.~Hanna, H.~Wen, J.~A. Retamero, W.~A. Moye, R.~Yousfi, C.~Kanan, D.~S. Klimstra, B.~Rothrock, S.~Liu, T.~J. Fuchs, A foundation model for clinical-grade computational pathology and rare cancers detection, Nature Medicine (2024) 1--12.

\bibitem{Zimmermann2024}
E.~Zimmermann, E.~Vorontsov, J.~Viret, A.~Casson, M.~Zelechowski, G.~Shaikovski, N.~Tenenholtz, J.~Hall, D.~Klimstra, R.~Yousfi, T.~Fuchs, N.~Fusi, S.~Liu, K.~Severson, Virchow 2: Scaling self-supervised mixed magnification models in pathology, arXiv preprint arXiv:2408.00738 (2024).

\bibitem{hoptimus0}
C.~Saillard, R.~Jenatton, F.~Llinares-López, Z.~Mariet, D.~Cahané, E.~Durand, J.-P. Vert, \href{https://github.com/bioptimus/releases/tree/main/models/h-optimus/v0}{H-optimus-0} (2024).
\newline\urlprefix\url{https://github.com/bioptimus/releases/tree/main/models/h-optimus/v0}

\bibitem{Achiam2023}
{OpenAI}, J.~Achiam, S.~Adler, S.~Agarwal, L.~Ahmad, I.~Akkaya, F.~L. Aleman, D.~Almeida, J.~Altenschmidt, S.~Altman, S.~Anadkat, R.~Avila, I.~Babuschkin, S.~Balaji, V.~Balcom, P.~Baltescu, H.~Bao, M.~Bavarian, J.~Belgum, I.~Bello, J.~Berdine, G.~Bernadett-Shapiro, C.~Berner, L.~Bogdonoff, O.~Boiko, M.~Boyd, A.-L. Brakman, G.~Brockman, T.~Brooks, M.~Brundage, K.~Button, T.~Cai, R.~Campbell, A.~Cann, B.~Carey, C.~Carlson, R.~Carmichael, B.~Chan, C.~Chang, F.~Chantzis, D.~Chen, S.~Chen, R.~Chen, J.~Chen, M.~Chen, B.~Chess, C.~Cho, C.~Chu, H.~W. Chung, D.~Cummings, J.~Currier, Y.~Dai, C.~Decareaux, T.~Degry, N.~Deutsch, D.~Deville, A.~Dhar, D.~Dohan, S.~Dowling, S.~Dunning, A.~Ecoffet, A.~Eleti, T.~Eloundou, D.~Farhi, L.~Fedus, N.~Felix, S.~P. Fishman, J.~Forte, I.~Fulford, L.~Gao, E.~Georges, C.~Gibson, V.~Goel, T.~Gogineni, G.~Goh, R.~Gontijo-Lopes, J.~Gordon, M.~Grafstein, S.~Gray, R.~Greene, J.~Gross, S.~S. Gu, Y.~Guo, C.~Hallacy, J.~Han, J.~Harris, Y.~He, M.~Heaton, J.~Heidecke, C.~Hesse, A.~Hickey,
  W.~Hickey, P.~Hoeschele, B.~Houghton, K.~Hsu, S.~Hu, X.~Hu, J.~Huizinga, S.~Jain, S.~Jain, J.~Jang, A.~Jiang, R.~Jiang, H.~Jin, D.~Jin, S.~Jomoto, B.~Jonn, H.~Jun, T.~Kaftan, Łukasz Kaiser, A.~Kamali, I.~Kanitscheider, N.~S. Keskar, T.~Khan, L.~Kilpatrick, J.~W. Kim, C.~Kim, Y.~Kim, J.~H. Kirchner, J.~Kiros, M.~Knight, D.~Kokotajlo, Łukasz Kondraciuk, A.~Kondrich, A.~Konstantinidis, K.~Kosic, G.~Krueger, V.~Kuo, M.~Lampe, I.~Lan, T.~Lee, J.~Leike, J.~Leung, D.~Levy, C.~M. Li, R.~Lim, M.~Lin, S.~Lin, M.~Litwin, T.~Lopez, R.~Lowe, P.~Lue, A.~Makanju, K.~Malfacini, S.~Manning, T.~Markov, Y.~Markovski, B.~Martin, K.~Mayer, A.~Mayne, B.~McGrew, S.~M. McKinney, C.~McLeavey, P.~McMillan, J.~McNeil, D.~Medina, A.~Mehta, J.~Menick, L.~Metz, A.~Mishchenko, P.~Mishkin, V.~Monaco, E.~Morikawa, D.~Mossing, T.~Mu, M.~Murati, O.~Murk, D.~Mély, A.~Nair, R.~Nakano, R.~Nayak, A.~Neelakantan, R.~Ngo, H.~Noh, L.~Ouyang, C.~O'Keefe, J.~Pachocki, A.~Paino, J.~Palermo, A.~Pantuliano, G.~Parascandolo, J.~Parish, E.~Parparita,
  A.~Passos, M.~Pavlov, A.~Peng, A.~Perelman, F.~de~Avila Belbute~Peres, M.~Petrov, H.~P. de~Oliveira~Pinto, Michael, Pokorny, M.~Pokrass, V.~H. Pong, T.~Powell, A.~Power, B.~Power, E.~Proehl, R.~Puri, A.~Radford, J.~Rae, A.~Ramesh, C.~Raymond, F.~Real, K.~Rimbach, C.~Ross, B.~Rotsted, H.~Roussez, N.~Ryder, M.~Saltarelli, T.~Sanders, S.~Santurkar, G.~Sastry, H.~Schmidt, D.~Schnurr, J.~Schulman, D.~Selsam, K.~Sheppard, T.~Sherbakov, J.~Shieh, S.~Shoker, P.~Shyam, S.~Sidor, E.~Sigler, M.~Simens, J.~Sitkin, K.~Slama, I.~Sohl, B.~Sokolowsky, Y.~Song, N.~Staudacher, F.~P. Such, N.~Summers, I.~Sutskever, J.~Tang, N.~Tezak, M.~B. Thompson, P.~Tillet, A.~Tootoonchian, E.~Tseng, P.~Tuggle, N.~Turley, J.~Tworek, J.~F.~C. Uribe, A.~Vallone, A.~Vijayvergiya, C.~Voss, C.~Wainwright, J.~J. Wang, A.~Wang, B.~Wang, J.~Ward, J.~Wei, C.~Weinmann, A.~Welihinda, P.~Welinder, J.~Weng, L.~Weng, M.~Wiethoff, D.~Willner, C.~Winter, S.~Wolrich, H.~Wong, L.~Workman, S.~Wu, J.~Wu, M.~Wu, K.~Xiao, T.~Xu, S.~Yoo, K.~Yu, Q.~Yuan,
  W.~Zaremba, R.~Zellers, C.~Zhang, M.~Zhang, S.~Zhao, T.~Zheng, J.~Zhuang, W.~Zhuk, B.~Zoph, Gpt-4 technical report, arXiv preprint arXiv:2303.08774 (2023).

\bibitem{Touvron2023}
H.~Touvron, L.~Martin, K.~Stone, P.~Albert, A.~Almahairi, Y.~Babaei, N.~Bashlykov, S.~Batra, P.~Bhargava, S.~Bhosale, D.~Bikel, L.~Blecher, C.~C. Ferrer, M.~Chen, G.~Cucurull, D.~Esiobu, J.~Fernandes, J.~Fu, W.~Fu, B.~Fuller, C.~Gao, V.~Goswami, N.~Goyal, A.~Hartshorn, S.~Hosseini, R.~Hou, H.~Inan, M.~Kardas, V.~Kerkez, M.~Khabsa, I.~Kloumann, A.~Korenev, P.~S. Koura, M.-A. Lachaux, T.~Lavril, J.~Lee, D.~Liskovich, Y.~Lu, Y.~Mao, X.~Martinet, T.~Mihaylov, P.~Mishra, I.~Molybog, Y.~Nie, A.~Poulton, J.~Reizenstein, R.~Rungta, K.~Saladi, A.~Schelten, R.~Silva, E.~M. Smith, R.~Subramanian, X.~E. Tan, B.~Tang, R.~Taylor, A.~Williams, J.~X. Kuan, P.~Xu, Z.~Yan, I.~Zarov, Y.~Zhang, A.~Fan, M.~Kambadur, S.~Narang, A.~Rodriguez, R.~Stojnic, S.~Edunov, T.~Scialom, Llama 2: Open foundation and fine-tuned chat models, arXiv preprint arXiv:2307.09288 (2023).

\bibitem{Campanella2024}
G.~Campanella, S.~Chen, R.~Verma, J.~Zeng, A.~Stock, M.~Croken, B.~Veremis, A.~Elmas, K.~lin Huang, R.~Kwan, J.~Houldsworth, A.~J. Schoenfeld, C.~Vanderbilt, A clinical benchmark of public self-supervised pathology foundation models, arXiv preprint arXiv:2407.06508 (2024).

\bibitem{Chen2024}
R.~J. Chen, T.~Ding, M.~Y. Lu, D.~F.~K. Williamson, G.~Jaume, A.~H. Song, B.~Chen, A.~Zhang, D.~Shao, M.~Shaban, M.~Williams, L.~Oldenburg, L.~L. Weishaupt, J.~J. Wang, A.~Vaidya, L.~P. Le, G.~Gerber, S.~Sahai, W.~Williams, F.~Mahmood, Towards a general-purpose foundation model for computational pathology, Nature Medicine (2024) 1--13.

\bibitem{Neidlinger2024}
P.~Neidlinger, O.~S. M.~E. Nahhas, H.~S. Muti, T.~Lenz, M.~Hoffmeister, H.~Brenner, M.~van Treeck, R.~Langer, B.~Dislich, H.~M. Behrens, C.~Röcken, S.~Foersch, D.~Truhn, A.~Marra, O.~L. Saldanha, J.~N. Kather, Benchmarking foundation models as feature extractors for weakly-supervised computational pathology, arXiv preprint arXiv:2408.15823 (2024).

\bibitem{Asadi2024ocean}
M.~Asadi-Aghbolaghi, H.~Farahani, A.~Zhang, A.~Akbari, S.~Kim, A.~Chow, S.~Dane, {OCEAN Challenge Consortium}, {OTTA Consortium}, D.~G. Huntsman, C.~B. Gilks, S.~Ramus, M.~K{\"o}bel, A.~N. Karnezis, A.~Bashashati, Machine learning-driven histotype diagnosis of ovarian carcinoma: Insights from the ocean ai challenge, medRxiv (2024) 2024--04.

\bibitem{Allen2023}
K.~E. Allen, J.~Breen, G.~Hall, K.~Zucker, N.~Ravikumar, N.~M. Orsi, $\#$900 comparative evaluation of ovarian carcinoma subtyping in primary versus interval debulking surgery specimen whole slide images using artificial intelligence, International Journal of Gynecologic Cancer 33~(Suppl 3) (2023) A429--A430.
\newblock \href {https://doi.org/10.1136/ijgc-2023-ESGO.900} {\path{doi:10.1136/ijgc-2023-ESGO.900}}.

\bibitem{Ilse2018}
M.~Ilse, J.~Tomczak, M.~Welling, Attention-based deep multiple instance learning, in: International conference on machine learning, PMLR, 2018, pp. 2127--2136.

\bibitem{Song2023}
A.~H. Song, G.~Jaume, D.~F. Williamson, M.~Y. Lu, A.~Vaidya, T.~R. Miller, F.~Mahmood, Artificial intelligence for digital and computational pathology, Nature Reviews Bioengineering 1~(12) (2023) 930--949.

\bibitem{Dosovitskiy2020}
A.~Dosovitskiy, L.~Beyer, A.~Kolesnikov, D.~Weissenborn, X.~Zhai, T.~Unterthiner, M.~Dehghani, M.~Minderer, G.~Heigold, S.~Gelly, J.~Uszkoreit, N.~Houlsby, An image is worth 16x16 words: Transformers for image recognition at scale, arXiv preprint arXiv:2010.11929 (2020).

\bibitem{Aben2024}
{kaiko.ai}, N.~Aben, E.~D. de~Jong, I.~Gatopoulos, N.~Känzig, M.~Karasikov, A.~Lagré, R.~Moser, J.~van Doorn, F.~Tang, Towards large-scale training of pathology foundation models, arXiv preprint arXiv:2404.15217 (2024).

\bibitem{Oquab2023}
M.~Oquab, T.~Darcet, T.~Moutakanni, H.~Vo, M.~Szafraniec, V.~Khalidov, P.~Fernandez, D.~Haziza, F.~Massa, A.~El-Nouby, et~al., Dinov2: Learning robust visual features without supervision, arXiv preprint arXiv:2304.07193 (2023).

\bibitem{Reinhard2001}
E.~Reinhard, M.~Adhikhmin, B.~Gooch, P.~Shirley, Color transfer between images, IEEE Computer graphics and applications 21~(5) (2001) 34--41.

\bibitem{Macenko2009}
M.~Macenko, M.~Niethammer, J.~S. Marron, D.~Borland, J.~T. Woosley, X.~Guan, C.~Schmitt, N.~E. Thomas, A method for normalizing histology slides for quantitative analysis, in: 2009 IEEE international symposium on biomedical imaging: from nano to macro, IEEE, 2009, pp. 1107--1110.

\bibitem{Otsu1975}
N.~Otsu, A threshold selection method from gray-level histograms, Automatica 11~(285-296) (1975) 23--27.

\bibitem{Kingma2014}
D.~P. Kingma, J.~Ba, Adam: A method for stochastic optimization, arXiv preprint arXiv:1412.6980 (2014).

\bibitem{Benjamini1995}
Y.~Benjamini, Y.~Hochberg, Controlling the false discovery rate: a practical and powerful approach to multiple testing, Journal of the Royal statistical society: series B (Methodological) 57~(1) (1995) 289--300.

\bibitem{Collins2024}
G.~S. Collins, K.~G.~M. Moons, P.~Dhiman, R.~D. Riley, A.~L. Beam, B.~Van~Calster, M.~Ghassemi, X.~Liu, J.~B. Reitsma, M.~van Smeden, A.-L. Boulesteix, J.~C. Camaradou, L.~A. Celi, S.~Denaxas, A.~K. Denniston, B.~Glocker, R.~M. Golub, H.~Harvey, G.~Heinze, M.~M. Hoffman, A.~P. Kengne, E.~Lam, N.~Lee, E.~W. Loder, L.~Maier-Hein, B.~A. Mateen, M.~D. McCradden, L.~Oakden-Rayner, J.~Ordish, R.~Parnell, S.~Rose, K.~Singh, L.~Wynants, P.~Logullo, Tripod+ ai statement: updated guidance for reporting clinical prediction models that use regression or machine learning methods, bmj 385 (2024).

\bibitem{Bibal2022}
A.~Bibal, R.~Cardon, D.~Alfter, R.~Wilkens, X.~Wang, T.~Fran{\c{c}}ois, P.~Watrin, Is attention explanation? an introduction to the debate, in: Proceedings of the 60th Annual Meeting of the Association for Computational Linguistics (Volume 1: Long Papers), 2022, pp. 3889--3900.

\bibitem{Janowczyk2019}
A.~Janowczyk, R.~Zuo, H.~Gilmore, M.~Feldman, A.~Madabhushi, Histoqc: an open-source quality control tool for digital pathology slides, JCO clinical cancer informatics 3 (2019) 1--7.

\bibitem{Shakhawat2023}
H.~Shakhawat, S.~Hossain, A.~Kabir, S.~H. Mahmud, M.~M. Islam, F.~Tariq, Review of artifact detection methods for automated analysis and diagnosis in digital pathology, in: Artificial Intelligence For Disease Diagnosis And Prognosis In Smart Healthcare, CRC Press, 2023, pp. 177--202.

\bibitem{Breen2024}
J.~Breen, K.~Zucker, K.~Allen, N.~Ravikumar, N.~M. Orsi, Generative adversarial networks for stain normalisation in histopathology, in: Applications of Generative AI, Springer International Publishing, Cham, 2024, pp. 227--247.
\newblock \href {https://doi.org/10.1007/978-3-031-46238-2\_11} {\path{doi:10.1007/978-3-031-46238-2\_11}}.

\bibitem{Dooper2023}
S.~Dooper, H.~Pinckaers, W.~Aswolinskiy, K.~Hebeda, S.~Jarkman, J.~van~der Laak, G.~Litjens, {BIGPICTURE Consortium}, Gigapixel end-to-end training using streaming and attention, Medical Image Analysis 88 (2023) 102881.

\bibitem{Shao2023}
Z.~Shao, L.~Dai, Y.~Wang, H.~Wang, Y.~Zhang, Augdiff: Diffusion based feature augmentation for multiple instance learning in whole slide image, arXiv preprint arXiv:2303.06371 (2023).

\bibitem{Wang2020}
Y.~Wang, D.~Farnell, H.~Farahani, M.~Nursey, B.~Tessier-Cloutier, S.~J. Jones, D.~G. Huntsman, C.~B. Gilks, A.~Bashashati, Classification of epithelial ovarian carcinoma whole-slide pathology images using deep transfer learning, arXiv preprint arXiv:2005.10957 (2020).

\end{thebibliography}
\end{document}